\documentclass[12pt]{article}
\usepackage{latexsym}
\usepackage{amssymb}
\usepackage{amsmath}
\usepackage{graphicx}
\usepackage{float}
\usepackage{amsthm}

\usepackage{xcolor}

\usepackage[all]{xy}

\newtheorem{theorem}{Theorem}[section]
\newtheorem{prop}[theorem]{Proposition}
\newtheorem{notn}[theorem]{Notation}

\newtheorem{cor}[theorem]{Corollary}
\newtheorem{definition}[theorem]{Definition}
\newtheorem{remark}[theorem]{Remark}

\newcommand{\mn}{\medskip\noindent}
\newcommand{\RR}{\mathbb R}
\newcommand{\FF}{\mathcal{F}}
\newcommand{\p}{\partial}
\newcommand{\ve}{\varepsilon}
\newcommand{\tq}{\widetilde{q_2}}
\newcommand{\tqq}{\widetilde{q_3}}
\title{Bitangent planes of surfaces and applications to thermodynamics}
\author{Peter Giblin \and Graham Reeve}
\begin{document}
\maketitle

\abstract {   The classical van der Waals equation,  applied to one or two
mixing fluids, and the Helmholtz (free) energy function $A$ yield,
for fixed temperature $T$, a curve in the plane
$\RR^2$ (one fluid) or a surface in 3-space $\RR^3$ (binary fluid). A line tangent to this curve in two
places (bitangent line), or a set of planes tangent to this surface in two places 
(bitangent planes) have a thermodynamic
significance which is well documented in the classical literature. Points of contact of
bitangent planes trace `binodal curves' on the surface in $\RR^3$. The study of these bitangents is
also classical, starting  with D.J. Korteweg and J.D.
van der Waals at the end of the $19^{\rm th}$ 
century, but continuing into modern times. In this paper we give a summary of
the thermodynamic background and of other mathematical investigations and then
present a new mathematical approach which classifies a wide range of
situations in $\RR^3$ where bitangents occur.  In particular, we are able
to justify many of the details in diagrams of binodal curves
observed by Korteweg and
others, using techniques from singularity theory.
}

\medskip\noindent
MR2020: 58K05; 53A05, 53B50, 80A10.

\medskip\noindent
Key words: curve, surface, bitangent, contact equivalence, normal form,
criminant, binodal, Helmholtz free energy, van der Waals equation, binary fluid, isotherm.

\section{Introduction}\label{s:intro}

The origins of this work are twofold. Firstly, on the mathematical side,
Giblin and Zakalyukin (\cite{GZ1,GZ2}) have studied
so-called {\em centre symmetry set (CSS)} of a pair of smooth surfaces in real euclidean 3-space
(or two disjoint regions on a single surface). 
A part of this work (\cite[\S4]{GZ1},\cite[\S3]{GZ2}) concerns two local surface patches $M, N$ having a common tangent plane:
for the CSS we consider all pairs of {\em parallel} tangent planes to $M$ and $N$ and the
`chords' joining their points of contact. This 2-parameter family of chords has, in suitable circumstances a real local
envelope, that is a real surface $S$ tangent to all the chords (regarded here as infinite straight lines). The generally singular surface $S$ is the CSS of $M$ and $N$.
Within this family of chords there may be a 1-parameter family for which the corresponding
parallel planes actually coincide---we say the plane is a `bitangent plane'---and the corresponding 
chords play a special role within the CSS.  The contact points of the bitangent
planes trace {\em binodal curves}  \cite{Meijer} on $M$ and $N$ (also called {\em coexistence curves}
\cite{Callen} and 
{\em conodal curves}  or {\em connodal curves} \cite{Ricardo,Sengers} ).
The local structure of these curves will
depend on the geometrical properties of $M$ and $N$. The straight line joining 
points of contact of bitangent planes is called a {\em tie line} or {\em 
bitangent chord}, and these lines form a ruled surface.

In part
we are responding
to a suggestion on page 85 of \cite{Sengers}:
`It would be interesting to compare
Korteweg's method of continuous
deformation of surfaces with the
methodology of catastrophe theory'. But the major part of our
mathematical contribution is to give a complete list of  normal
forms of singularities up to codimension 1 (generic 1-parameter
families of surfaces)  occurring in the bitangent plane context, extending the lists in
\cite{GZ1} and \cite{RZ}. 

Secondly, the work of D.J.Korteweg on thermodynamic stability of
mixtures of two fluids, as recounted in \cite{Sengers}, 
necessitates a study of bitangent planes
of surfaces. For Korteweg, the surface
in question is the isothermal surface defined by the `Helmholtz (free) energy function' $A(V,x)$. In Section~\ref{s:thermo}
we shall briefly describe this function, and give some details of the role that bitangent
planes play, first describing the simpler situation where a surface is replaced by a
plane curve and bitangent planes by bitangent lines. References for
Helmholtz free energy include \cite{Callen,Denbigh}.

\medskip

In this article we shall not explicitly study thermodynamic stability or the
consequent need for surfaces which are convex (see for example \cite{RS})
but instead place the study of binodal curves and related surface geometry in
the context of modern singularity theory, combining geometrical information 
which is invariant to affine transformations of 3-space with more
qualitative results invariant to local diffeomorphisms of the ambient space. The
latter results are obtained by reducing families of functions to normal forms
using appropriate equivalence relations---these preserve essential structure such as cusps, but
do not necessarily preserve all geometrical features. In fact it is the ability to use
appropriate equivalence relations on families of functions, and then to invoke
the tools of singularity theory to reduce to normal forms, which distinguishes
our work from that of Korteweg, and also from the work of the authors 
mentioned in \S\ref{s:other}.  From a singularity point of view, this work also extends the work of \cite{GZ1, GZ2} to include a classification of centre symmetry sets for 1-parameter of surfaces in 3-space in the vicinity of one the surfaces.

In more detail, the structure of this article is as follows.  In \S\ref{s:other} we briefly
describe some other work in the same area as our article. 
In \S\ref{s:thermo} we give a sketch of the thermodynamic background,
starting with the simpler situation of a single fluid and continuing
with the two fluid case. In \S\ref{s:setup} we
give two mathematical approaches to the study of binodal curves, the first a direct
approach which allows us to describe the local
geometry of the surface patches in the various
cases, and the second a more general method via
`generating functions' which leads to normal forms
and in addition to a more precise description of singularities for generic 1-parameter families of surfaces. The local geometry is further explained 
in \S\ref{ss:geom}. In \S\ref{ss:representation}
we use the normal forms method to make pictures of
the binodal curves and of the ruled surface of
tie lines which preserve singularities up
to local diffeomorphism, but not the local
geometry of the surfaces and curves such as
curvature and inflexions. Calculations and sample
proofs are in \S\ref{s:proofs} and we make some
concluding remarks in \S\ref{s:conclusions}.

\medskip

We restrict ourselves in this article to the `bilocal' case, that is bitangent planes
having contact with two separated surface patches, which could nevertheless be  part of the same larger surface. The `local' case, typified by 
bitangent planes both of whose contact points are in an arbitrarily small neighbourhood of
a cusp of Gauss (plait point, godron) on a single surface patch, gives a number of additional cases and we
hope to cover these elsewhere.

\section{Other work in this area}\label{s:other}
A.N.Varchenko in \cite{Varchenko}  considers a
different thermodynamic potential from the
Helmholtz potential, one which is
a function of pressure $P,$ temperature $T$ and proportion of two fluids $x$. (The main results are for a 2-component
mixture with a single variable $x$, but there is also a general treatment for larger numbers of components.)
For a fixed pressure and temperature the 
thermodynamic potential of each homogeneous phase
has its own graph as a function of $x$; these graphs a convex downwards and
generally intersect. For a given $x$ thermodynamic
stability requires that the actual value of the potential follows the convex hull
of the  homogeneous phase graphs. Varchenko therefore studies the
evolution of these convex hulls of graphs in the plane,
varying in a 2-parameter family (parameters $P$ and $T$). 

P.H.E.Meijer in \cite{Meijer} investigates the evolution of tie lines and  binodal
curves by means of differential equations, and his article contains a clear
statement of the underlying physics. In particular he identifies the evolution of
binodal curves which we call `lips' in \S\ref{ss:geom} (he gives the typical `lips' diagram the equally
descriptive name `Napoleon's hat').

Ricardo Uribe-Vargas in \cite{Ricardo} undertakes a detailed investigation of the
behaviour of binodal curves close to cusps of Gauss 
and their relationship with other curves such as the parabolic curve. In this case
the binodal and the parabolic curves are tangent at the cusp of Gauss and this
creates a region between them 
in which the surface is convex but not thermodynamically
stable on account of the double tangencies. The author also studies
the evolution of these curves, and others,  during deformations of the surface.

In an unpublished Master's thesis \cite{Olsen} (see also \cite{DO}), W.E.Olsen, a student of Daniel Dreibelbis,
studies bitangent lines and planes to two surface patches $M,N$ by considering the
corresponding subsets of $M\times N$ and finds geometrical conditions for
singularities of  their projections to the two factors.
He also gives examples of 1-parameter deformations of the surfaces and considers
the more degenerate case of double-points, that is points of intersection of $M$ and $N$.

Determination of plait points (cusps of Gauss, godrons) 
on surfaces connected with thermodynamic equilibrium continues to be
of practical interest; see for example~\cite{MSL}. 

 The situation considered in this article, that is where we consider singularities in the vicinity of one of the surfaces, is also called `on shell', and various applications for this have been explored including for example in semiclassical physics, see \cite{Wojtek, Wojtek2}.

\section{Thermodynamic background}\label{s:thermo}
\subsection{One fluid}\label{ss:1fluid}
\begin{figure}[h]
\centering
\includegraphics[width=0.5\textwidth]{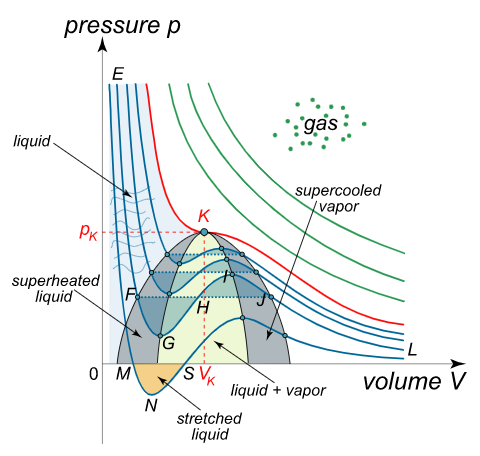}
\caption{\small Van der Waals isotherms in the $PV$-plane (reproduced with permission from math24.net \cite{math24}). (The outer boundary through $F,K,J$ is
the binodal  curve and  the inner boundary through $G,K,I$ is the spinodal curve.)}
\label{PV-diagram}
\end{figure}

For a fixed temperature $T$, the Van der Waals equation 
\begin{equation}
\left(P+\frac{a}{V^2}\right)(V-b)=RT,
\label{eq:vdW}
\end{equation}
where $a,b$ are constants depending on the substance and $R$ is an absolute
constant,
describes the relation between pressure $P$ and molar volume $V$ of a single fluid. In the $(P,V)$-plane, this dependence is represented as a family of isotherms, that is curves of constant temperature
$T$; see Figure \ref{PV-diagram}.

For a given system, if we change the volume, or the pressure, of the fluid under some isothermal process (where the temperature stays constant) the pressure will be given by following one of these isotherms.  Typically an isothermal process occurs when a system is in contact with an outside thermal reservoir, and a change in the system occurs slowly enough to allow the system to be continuously adjusted to the temperature of the reservoir through heat exchange.

For temperatures below the so called critical temperature (the red curve through $K$ in Figure \ref{PV-diagram}), the isotherms have an undulating shape. A portion of such an isotherm between its local minimum and local maximum has a positive derivative $P_V > 0$, which corresponds to an {\em unstable} state of the substance. (Here and
below we use suffices to denote partial derivatives.) Any small positive perturbation of the pressure $dP >0$ causes an increase in the fluid volume (since $dV > 0$), which in theory would lead to an explosive expansion of the fluid. If the curve slopes downwards, that is $P_V < 0$, then increasing the pressure would result in a  decrease in volume ($dV<0)$.  In reality, a liquid-gas phase transition occurs in this part of the $PV$-diagram. This transition is accompanied by a significant change in the volume $V$ at constant pressure $P$ and constant temperature $T$.  Such a transition is represented by a straight horizontal line, called a {\em tie line}, in the $PV$-diagram (the segment line $FJ$  in Figure  \ref{PV-diagram}). 
For example, an isothermal process would follow the tie line $FHJ$, rather than $FGHIJ$, along which the two phases liquid/vapour coexist in varying proportions in the container at this moment.

The location of the horizontal section $FHJ$ is determined from thermodynamic considerations,
resulting in the so-called Maxwell equal area rule. This states that the areas of the curved shapes $FGH$ and $HIJ$ should be equal (see \cite{Callen}, page 238 for details).

The set of start and end points of the horizontal segment lines at different temperatures form a curve, which is called the {\em binodal} curve. The curve passing through the local minima and maxima of all the isotherms is called the {\em spinodal} curve. 
The unstable states (for which $P_V > 0$) are within the region bounded by the spinodal curve. The part of the diagram between the spinodal and binodal curves in principle satisfies the stability criterion $P_V < 0$; the states in this region are called {\em metastable} states. The left half of the indicated region represents the superheated liquid, and the right half corresponds to the supercooled vapour.  Superheated liquid is familiar to anyone who has heated a liquid in a microwave to above boiling point, only to have it suddenly boil after inserting a spoon or some other nucleation site.  

\begin{figure}[H]
\begin{center}
\hspace*{-0.5in}
\includegraphics[width=5in]{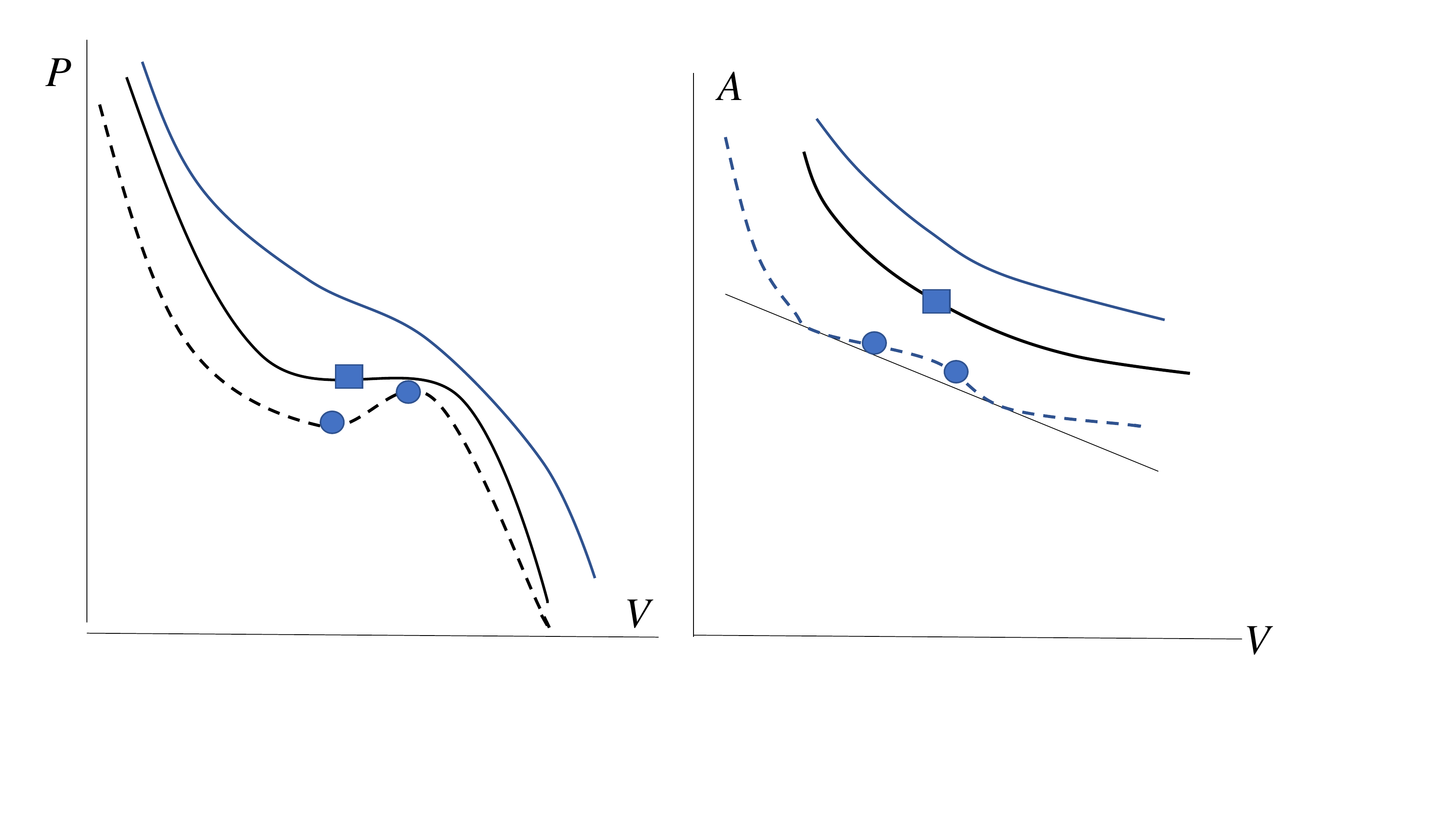}
\end{center}
\vspace{-0.7in}
\caption{\small The $PV$ curve versus the $AV$ curve where $A$ is the Helmholtz
energy, $A_V=-P$, for three values of the temperature, increasing from top to bottom. The middle curve is the critical $PV$ curve, with a horizontal inflexion (at the square), corresponding to an `undulation' 
on the $AV$ curve---a `very flat' point where the tangent line has 4-point contact with the
curve.  Decreasing the temperature `resolves' this undulation to a curve with
increasing (negative) gradient and increasing temperature  to a curve with two inflexions, marked by dots.
This permits the formation of a double tangent. All the $AV$ curves must be 
decreasing as $V$ increases, since $P>0$. }
\label{fig:AV}
\end{figure}

The Helmholtz  energy $A$  is a {\em thermodynamic potential}
 that measures the useful work obtainable from a closed thermodynamic system.
 For a single fluid it is defined by 
 \begin{equation}\label{Aplanecase}
 A=-RT\ln(V-b)-\frac{a}{V^2} \mbox{ so that } A_V=-\frac{RT}{V-b}+\frac{a}{V^2} = -P
 \end{equation}
 from Van der Waal's equation (\ref{eq:vdW}).  See Figure~\ref{fig:AV} and \S\ref{ss:2fluids}
 below for further discussion.  The existence of maxima and
 minima on the $PV$ curve implies inflexions, given by $A_{VV}=0$
 on the $AV$ curve and hence a double tangent. In fact the conditions on the
 contact points of the double tangent line can be interpreted as saying the
 pressure $P$ and the `thermodynamic potential' $\mu$ are the same at these points.
 We give more details of this in \S\ref{ss:2fluids}.
 
\subsection{Binary fluids}\label{ss:2fluids}
Thermodynamic theory has a concept of  `internal energy'  $E$, which is a function of the `extensive variables' $S,V,N_1,...N_r$.  Here $S$ is entropy, $V$ is molar volume  and the $N_i$ are measures of the molar quantities of the
substances present.  From this function other `intensive variables' are determined,
in fact these can be regarded as definitions:
Temperature $T=E_S$, pressure $P=-E_V$ and `chemical potentials' $\mu_i= 
E_{N_i}$.  Let us assume the
$N_i$ are constant, so that $E$ is a function of $S$ and $V$. By means of 
`Legendre transformations'---essentially representing surfaces by their duals---$E$
can be regarded, at any rate locally,  as a function of any pair of variables among $S,V,P,T$.
Helmholtz  energy $A$  is defined in this context by $A=E-TS$ and it measures
the amount of energy needed to create a system in the absence of changes in
temperature and volume, when account is taken of spontaneous transfer
of energy from the environment.

In general, given a function $f(x,y)$ where $x=x(a,b)$ and $y=y(a,b)$ then
the chain rule says, for example, that
$f_b|a=f_x|y\times x_b|a+f_y|x \times y_b|a$, where $f_b|a$ for example says
that $f$ is regarded as a function of $a$ and $b$ and $a$ is held constant in
the differentiation with respect to $b$. Applying this to $f=E, x=S, y=V, a=T$ and $b=V$ gives, after
rearrangement and using $E_S|V=T$,
$ E_V|S=E_V|T-T S_V|T$.  We then obtain the fact that the Helmholtz energy $A=E-TS$
has the important property
\begin{equation}
-P=E_V|S=(E-TS)_V|T=A_V|T. 
\label{eq:A_VconstT}
\end{equation}

On the other hand we can find $P$ from the Van der Waals equation
(\ref{eq:vdW}) which, for two fluids in the proportion $x:1-x$, becomes
\[ \left(P+\frac{a(x)}{V^2}\right)(V-b(x))=RT,\]
where $a$ and $b$ now depend on $x$. Keeping $x$ fixed and solving this for $P$ 
then integrating with respect to  $V$  gives 
\[P=\frac{RT}{V-b(x)}-\frac{a(x)}{V^2}, \ \ \ A=-RT \ln(V - b(x)) -\frac{a(x)}{V}+\mbox{ function of } T.\]
This function of $T$ is known as the `entropy mixing term' and for two fluids it
is \\ $-R(x \ln x + (1 - x) \ln(1 - x))$ (see for example \cite[p.432]{Denbigh}). It is
also called the `Gibbs mixing term' in \cite[p.49]{Sengers}. Hence
\[A(V, x) = -RT \ln(V - b(x)) -\frac{a(x)}{V} + RT(x \ln x + (1 - x) \ln(1 - x)).\]
This is the family of isothermal (constant $T$)  surfaces in $(AVx)$-space which is the generalisation of the isothermal $AV$ curves
in Figure~\ref{fig:AV}. We recover the plane curve case (\ref{Aplanecase}) by substituting $x=1$,
remembering that $(1-x)\ln(1-x)\to 0$ as $x\to 1$. 

A normal vector to the isothermal surface at $u=(V_0,x_0,A(V_0,x_0))$ is
$(A_V( V_0,x_0), A_x(V_0,x_0),-1)$. Writing down the equation of the tangent plane at $u$ and substituting $V=x=0$ shows that the intercept of the tangent plane at $u$ with the
$A$-axis is $A(V_0,x_0)-V_0A_V(V_0,x_0)-x_0A_x(V_0,x_0)$. The condition for two tangent planes,
at $u_1$ and $u_2$ say, to be identical is therefore that $A_V(V_0,x_0), A_x(V_0,x_0)$ and
this intercept should coincide for $u=u_1$ and $u=u_2$. According to
(\cite[p.50]{Sengers}) this can be interpreted as saying that
the two  points $u_1$ and $u_2$  share the same pressure $P$ and the same chemical potentials 
$\mu_1,\mu_2$ of
the two components. 

For our mathematical investigation
 we are interested in studying the ruled surface of tie lines joining points of contact of
 bitangent planes (called the {\em criminant} surface below) to the surfaces $M$ and $N$ and reducing the configuration of
 $M$, $N$ and this ruled surface to `normal forms' under local diffeomorphism in order
 to examine its basic structure.

\section{The mathematical setup}\label{s:setup}
We consider two surface patches $M,N$ in $\RR^3$ with coordinates $(x,y,z)$, having
a common tangent plane $z=0$ at the points $(0,0,0)\in M, (0,1,0)\in N$. The
surfaces will be defined locally by $M: \{(u,v,f(u,v))\}; N: \{(x,1+y,g(x,y))\}$ where
$u,v,x,y$ are small and $f,g$ have no constant or degree 1 terms, that is they have zero 1-jets. We expand the functions $f,g$ by Taylor series about the base points $(0,0)$:
\begin{eqnarray}
f(u,v) &=& f_{20}u^2 + f_{11}uv + f_{02}v^2 + f_{30}u^3 + f_{21}u^2v + f_{12}uv^2 + f_{03}v^3 + + f_{40}u^4 + ... \nonumber  \\ 
g(x,y) &=& g_{20}x^2 + g_{11}xy + g_{02}y^2 + g_{30}x^3 + g_{21}x^2y + g_{12}xy^2 + g_{03}y^3  + ... 
\label{eq:local}
\end{eqnarray}
where the subscripts $ij$ indicate that the corresponding monomial is 
$u^iv^j$ or $x^iy^j$.

Sometimes we have to consider a {\em generic family of surfaces} in which case
the local parametrizations will vary according to a local parameter $\tau$ say, taken as close to 0.  In that
case we call the surfaces in the family $M_\tau$ and $N_\tau$ and write them locally
as 
\begin{eqnarray}
f(u,v, \tau) &=& f_{20}u^2 + f_{11}uv + f_{02}v^2 + f_{30}u^3 + f_{21}u^2v + f_{12}uv^2 + f_{03}v^3 + + f_{40}u^4 + ...  \nonumber \\ &+&  \tau(f_{031}v^3 + f_{121}uv^2 + ...) + \tau^2(f_{032}v^3 + f_{122}uv^2  + ... ) + ...\nonumber \\
g(x,y,\tau) &=& g_{20}x^2 + g_{11}xy + g_{02}y^2 + g_{30}x^3 + g_{21}x^2y + g_{12}xy^2 + g_{03}y^3  + ... \\ \nonumber &+& \tau(g_{101}x + g_{011}y+ g_{201}x^2 + g_{111}xy + g_{021}y^2 + g_{301}x^3 + ...) \\ \nonumber  &+&  \tau^2(g_{102}x + g_{012}y + g_{202}x^2 + g_{112}xy + ...) + ...
\label{eq:local-tau}
\end{eqnarray}
where the subscripts $ijk$ indicate that the corresponding monomial is $\tau^ku^iv^j$ or $\tau^kx^iy^j$, and for brevity, the subscripts $ij$ are used when $k$ is zero.  Here the
2-jet of $f$ has been retained by applying a smooth family of affine transformations.
A point of $M_\tau$ is $(u,v,f(u,v,\tau))$ and a point of $N_\tau$ is 
$(x,y+1,g(x,y,\tau))$. When $\tau=0$ we recover the base surfaces $M_0=M$ and
$N_0=N$ of the family.

\subsection{Equations defining binodal curves}\label{ss:equations}
To make explicit calculations of binodal curves using jets we can use the 
inverse image of $(0,0,0)$ under the map $H=(H_1,H_2,H_3):(\RR^4,0) \to (\RR^3,0)$
given by
\begin{equation}\label{eq:H}
 H(u,v,x,y)=(f_u-g_x, f_v-g_y, (x-u)f_u+(1+y-v)f_v+f-g),
\end{equation}
where subscripts representing {\em variables},
as opposed to subscripts representing
coefficients as in (\ref{eq:local})
and (\ref{eq:local-tau}), stand for partial derivatives. Functions are evaluated at
$(u,v)$ or $(x,y)$ as appropriate. Vanishing of the first two components of $H$ says that the
normal vectors to $M$ and $N$ are parallel and vanishing
of the third says that the line joining the
points $(u,v,f)$ and $(x,1+y,g)$ is perpendicular to the normal to $M$. Altogether
$H^{-1}({\bf 0})$ selects the points of $M$ and $N$ for which tangent planes
coincide. Projecting $H^{-1}({\bf 0})$ to the first two coordinates gives the binodal curve on $M$
and projecting to the last two coordinates gives the binodal curve on $N$. Studying these
projections can give geometrical information on the bionodal
 curves; an example
is given below and in \S\ref{ss:geom} we shall show how this direct method can be
used to obtain such information.

\begin{prop}\label{prop:K2}
Suppose that \\ {\rm (i)} The surface $N$ is parabolic at $(0,1,0)$, but does
not have a cusp of Gauss at that point. This amounts to saying that
the unique asymptotic direction at $N$ is not a root of the cubic terms of $g$, \\
{\rm (ii)} The unique asymptotic direction
for $N$ at $(0,1,0)$ is not along the tie line $x=0$, that is $x$ is not
a factor of the quadratic terms of $g$, \\
{\rm (iii)} The origin is not a
parabolic point on $M$, that is $f_{11}^2-2f_{20}f_{02}\ne 0$
in {\rm (\ref{eq:local})}.

\mn
Then the binodal curve on $M$ at the origin has an ordinary cusp. In addition, the limiting
tangent to this cusp is conjugate to the tie line.
\end{prop}
An ordinary cusp is locally diffeomorphic to $\{(t^2,t^3)\}$ for small $t$.
Writing the quadratic terms of $g$ as $(ax+by)^2$, (i) says that
$a^3g_{03} - a^2bg_{12} + ab^2g_{21} - b^3g_{30} \ne 0$ and (ii) says that
$b\ne 0$.

\begin{remark}\label{rem:K2}
{\rm
The conclusion of the proposition above is essentially Korteweg's  assertion called K2 in \cite[p.76]{Sengers};
proofs by Korteweg and others are in \cite{K1,K2,K3,K4,K5}.
}
\end{remark}
\medskip

The proof of Proposition~\ref{prop:K2} amounts to using the jets of $H$ as in (\ref{eq:H}) to solve for
$x$ and $y$ as functions of $u$. The 1-jet of the second
component $H_2$ of $H$ at
$u=v=x=y=0$ is 
\[ f_{11}u+2f_{02}v -2abx-b^2y\]
so that using $b\ne 0$
we can use $H_2=0$ to solve locally for $y=y(u,v,x)$.
We can then substitute for $y$ in $H_1=0$ and $H_3=0$;
these then have leading terms
in $u,v,x$ given by
\[ -(af_{11} - 2bf_{20})u/b - (2af_{02} - bf_{11})v/b, \mbox{ and }
 f_{11}u+2f_{02}v, \mbox{ respectively}.\]
 The second of these can be solved locally for $u(v,x)$ or $v(u,x)$ since $f_{11}, f_{02}$
 are not both zero. In either case we substitute in $H_1=0$ and solve for $u(x)$ and $v(x)$.  These have the form $u(x) = px^2+qx^3 +...$ and $v(x)=rx^2+sx^3$
 where 
 \[ps-qr=\frac{6(a^3g_{03} - a^2bg_{12} + ab^2g_{21} - b^3g_{30})}
 {b^6(f_{11}^2-4f_{20}f_{02})},\]
 which is nonzero by assumption. This proves that 
 the curve in $M$ (strictly the projection of this curve on to
 the tangent plane at the origin) has an ordinary cusp. 
 Finally the limiting tangent direction  $(p,r)$ 
 of the cusp is a nonzero multiple of $(-2f_{02},f_{11})$, 
 which is a vector conjugate to the tie line direction $(0,1)$ in the tangent plane.
 \hfill$\Box$.
 
 \subsection{Generating families}\label{ss:gen}
 The direct method above is hard to apply  to the situation where the
 surfaces $M$ and $N$ vary in a 1-parameter family $M_\tau, N_\tau$, though Korteweg, as
 reported in \cite{Sengers}, produced many interesting diagrams of the evolution
 of binodal curves and other surface features such as parabolic and flecnodal curves.
 We shall adopt here a more general approach, based on the idea of
 generating families.

Following the method used in \cite[p.44-45]{GZ2}, \cite{RZ}, 
we use the generating family, for each fixed $\tau$ near 0,
\[\FF(n,u,v,x,y,\lambda,q,\tau) = \langle \lambda(u,v, f(u,v,\tau)) + \mu 
(x,y+1,g(x,y,\tau))-q,n\rangle.\] 
Here, the angle brackets $\langle, \rangle$ denote inner product in 3-space,
$\lambda, \mu$ are barycentric coordinates on a real line (that is $\lambda+\mu=1$),
$n$ and $q$ are vectors in $\RR^3$ and $u,v,x,y,\tau$ are as above. For a single
generic surface the family parameter $\tau$ will be absent.

The {\em criminant} of $\FF$, for a fixed $\tau$, is the ruled surface  created by the tie lines and is
given by
\[ \Delta\FF=\left\{ q: \mbox {for some } (n,u,v,x,y,\lambda), 
(\FF=)\frac{\p\FF}{\p u}=\frac{\p\FF}{\p v}=\frac{\p\FF}{\p x}=\frac{\p\FF}{\p y}=\frac{\p\FF}{\p \lambda}=\frac{\p\FF}{\p n}=0 \right\}.\] 
In the definition of $\Delta\FF$ the term `$\FF=$' is bracketed since $\FF$ is automatically
equal to zero given the three equations $\p\FF/\p n=0$.
Suppose $\lambda\ne 0$ and $\lambda\ne 1$. Then $\Delta\FF$ is the set of points $q\in\RR^3$ which are on a straight line
joining a point of $M_\tau$ to a point of $N_\tau$ (from $\p\FF/\p n=0$), and the line joining these points lies
in a bitangent plane to $M_\tau$ and $N_\tau$ with common normal $n$ (from the other conditions). When $\lambda=0$, $q$ is a point of $N_\tau$ (and $n$ is parallel to
the normal to $N_\tau$ there and perpendicular to the line joining the two points of $M_\tau$ and $N\tau$);
likewise $\lambda=1$ gives points of $M$.  The surfaces $M_\tau$ and $N_\tau$  are usually called the 
{\em redundant components} of $\Delta\FF$. The closure of the part of
$\Delta\FF$ for $\lambda\ne 0,1$
consists exactly of the (infinite) straight lines joining
points which share a common tangent plane, that is the ruled surface containing
these lines.  The projection to $(u,v)$ or to $(x,y)$ gives the binodal curve in
$M_\tau$ or $N_\tau$ respectively.

\begin{remark}
{\rm
The  {\em caustic}
 $\Sigma\FF$ is the set of $q$ for which first partial derivatives with
respect to $u,v,x,y$ are 0 and the $4\times 4$ matrix of second 
partial derivatives is singular. The {\em centre symmetry set} is the union of $\Delta\FF$
and $\Sigma\FF$.
 It is discussed
in detail in \cite{GZ1,GZ2}.
}
\end{remark}

The point of
these definitions is that up to local diffeomorphism in $\RR^3$ the sets $\Delta\FF$ and $\Sigma\FF$ are
invariant under appropriate changes of coordinates (to be introduced  below), 
allowing us to reduce the
family $\FF$ to a normal form and deduce the local
structure from that. For geometrical information
such as conjugate and asymptotic directions we still need to make
explicit calculations, as in \S\ref{ss:equations}.

 We shall make use also
of {\em stabilisation} which means that, if a family
contains a nondegenenerate quadratic form in `extra'
variables not occurring elsewhere in the family, then this form
can be removed without affecting the
diffeomorphism type of the criminant (or the caustic). This is because the
zero partial derivatives with respect to those extra variables ensures that their
values are zero.

In order to study the binodal curves we need to work in a neighbourhood of
a point of $M$ or $N$, that is with $\lambda$ as above close to 1 or 0. In
principle any base value $\lambda=\lambda_0$  can be chosen and then we
write $\lambda=\lambda_0+\ve$. The `base point' on the second
coordinate axis in $\RR^3$
is then $(0,1-\lambda_0,0)$ and to work with small coordinates we write
$q=(q_1,q_2,q_3)=(q_1,\tq + 1-\lambda_0,q_3).$
For the following proposition we take $\lambda_0=0$
so that we are `working near to $N$' and $q=(q_1,\tq+1,q_3)$. Thus the overall
base point from which we expand our functions is given by
$x=y=u=v=0, q_1=\tq=q_3=0, \ve=0 \ (\lambda = 0),  n_1=n_2=0, n_3=1.$

\begin{prop}\label{reduction}
Using stabilisation, the family $\FF$ can be reduced, near to the surface $N$,
to the following family, where $\tq=q_2-1$, $q=(q_1,q_2,q_3)$.
\[ \Phi(u,v,\ve,q,\tau) = \ve f(u,v,\tau) + (1-\ve)g\left(\frac{q_1 -\ve u}{1-\ve}, \frac{\tq - \ve v + \ve}{ 1 - \ve},\tau \right)-q_3.\]
in variables $(u,v) \in \RR^2$ and parameters $\ve \in \RR$, $q \in \RR^3$ in a neighbourhood of $\ve = 0,q_1=\tq=q_3 = 0,\tau=0$.
\end{prop} 
\noindent
Proof. 
Writing the family $\mathcal{F}$ in the coordinate form we get \[\FF = An_1 + Bn_2 + Cn_3\] where \[A = \ve u + (1-\ve)x - q_1,\]  \[B = \ve v + (1-\ve)(y + 1)-\tq-1 \] and \[C = \ve f(u,v) + (1-\ve)g(x,y) - q_3\]
For $\ve $ small the functions $A$ and $B$ are regular with respect to $x$ and $y$ and so can be chosen as the coordinate functions instead of $x$ and $y$, that is
\[x = \frac{A+q_1 - \ve u}{1 - \ve}, \ \ \ \ y =\frac{B+\tq - \ve(v-1)}{1 - \ve}.\]
So in the new coordinates we have  $$\mathcal{F} = An_1 + Bn_2 + C(A,B,u,v,\ve,q)n_3$$ where the function $C$ does not depend on $n_1$ and $n_2$. Applying Hadamard’s lemma to the function $C$ we get
$$C(A,B,u,v,\ve,q) = C(0,0,u,v,\ve,q) + A\varphi_1 + B\varphi_2,$$
where $\varphi_1$ and $\varphi_2$ are smooth functions in $A,B,u,v,\ve,q$  which vanish at $A = B = \ve = q_1 = \tq  = q_3  = 0$. (This is because,
$g$ having no linear terms, $\p C/\p A = \p C /\p B=0$ at the base point.)

Now the function $\mathcal{F}$ takes the form $$\mathcal{F} = A(n_1 + \varphi_1n_3)+B(n_2 + \varphi_2n_3)+C(0,0,u,v,\ve,q)$$ where the first two terms represent a non degenerate quadratic form in the independent variables $A,(n_1 +\varphi_1n_3),B$ and $(n_2 +\varphi_2n_3)$. Therefore, the function $\mathcal{F}$ is stably-equivalent to the function $\Phi = C(0,0,u,v,\ve,q)$ being the restriction of the function $C$ to the subspace $A = B = 0 $. This completes the proof. \hfill$\Box$

\subsection{Space-time contact equivalence}
To study the local structure of the criminant (ruled tie line) surface in the vicinity of the surface, and hence also the local structure of the binodal curve, we reduce the generating function $\Phi$ to a normal form up to an appropriate equivalence relation which preserves the surface up to local diffeomorphism. The relevant equivalence relation is the following notion of space-time equivalence, adapted for 1-parameter families of surfaces from \cite{GZ1}, (see also \cite{GR}):

\begin{definition}\label{def:stc}
Two germs of families $F_1$ and $F_2$ with variables $u \in \mathbb{R}^2$,  (time) parameter $\lambda \in \mathbb{R}$, (space) parameter $q \in \mathbb{R}^3$, and (family) parameter $\tau \in \mathbb{R}$ are called space-time contact equivalent if there exists a nonzero function $\phi(u,\lambda,q, \tau)$ and a diffeomorphism ${\theta} : \mathbb{R}^2 \times \mathbb{R}^{1+3+1} \to \mathbb{R}^2 \times \mathbb{R}^{1+3+1}$, of the form 
\[{\theta}: (u,\lambda,q, \tau) \mapsto (U(u, \lambda,q, \tau),\Lambda(\lambda,q, \tau), Q(q, \tau), T(\tau))\] such that $\phi F_1 = F_2  \circ \theta $. \end{definition}

\begin{remark}\label{rem:geom}
{\rm
Space-time contact equivalence allows us to reduce a generating function
$\FF$ to one of a {\em finite} list of cases, each of which produces
a `model' of the criminant and the binodal curves. Thus we obtain
a finite list `diagrams' or `pictures' which represent the
essential features of the various cases which arise. This
representation does not, however, preserve all the geometrical
properties of the surfaces and curves involved. Cusps are preserved
on the binodal curves  but inflexions are not. To study inflexions
requires a different technique (`duals' of surfaces and curves)
and we hope to pursue this elsewhere. Surface singularies
such as cuspidal edges,
swallowtail points, Whitney umbrellas and the like are preserved.
This is what we mean by `essential features'.
There is more information about the geometry of the different
cases in \S\ref{ss:geom}, and about {\em drawing} the criminant
and binodal curves in \S\ref{ss:representation}.
}
\end{remark}

\subsection{Expanding the generating function}
We now proceed by expanding the generating function as a power series, and consider the lowest degree terms.  First redefine $q_3$ so that $\Phi$ (as in Proposition~\ref{reduction})
becomes divisible by $\ve$:
$\tqq=q_3-g(q_1,\tq,\tau)$, an allowable change of variable according to 
Definition~\ref{def:stc}.
We denote by $\Phi_0(u,v, \varepsilon, \tqq) = \Phi(u,v,\varepsilon,0,0,\tqq,0)$, the organising centre of the family, which has first few terms of the power series in $\varepsilon$ at the origin as
\begin{eqnarray*}
\Phi_0&=&  -q_3 + (f_{02}v^2 + f_{11}uv + f_{20}u^2 + f_{30}u^3 + ...)\varepsilon \\
&& + ( g_{02}  - 2g_{02}v - g_{11}u + g_{02}v^2 + g_{11}uv + g_{20}u^2 )\varepsilon^2 +  (g_{02}+g_{03} +...)\varepsilon^3 + ... 
\end{eqnarray*}
where the dots denote higher degree terms.

\begin{notn}\label{q}
From this point we shall revert to the notation $(q_1,q_2,q_3)$ for coordinates
in $\RR^3$, all of these quantities being understood to be close to 0.
\end{notn}

Consider the space $W$ of function germs of the type $\mathcal{F} = -q_3 +  \varepsilon \mathcal{H}(u, v,\varepsilon, q_1, q_2, \tau)$. Following \cite{GZ1,ReeveThesis} we show stability of the generating function inside this space $W$.  

The transversality theorem implies that since the base points of the two surfaces already share a bitangent plane (one condition) at $\mu_0=0$ only one extra conditions can be imposed on the derivatives of the surfaces.  It follows that there are three distinct generic singularity types that can occur on the surfaces.  In particular we shall prove the following propositions.

\begin{prop}\label{prop:2surfaces} {\rm (See Proposition~\ref{C3-1-proof}) } \ \ 
For a generic pair of surfaces $M$ and $N$ near $(0,1,0)\in N$ the  generating family germ $\Phi$ is space-time contact equivalent to one of the following normal forms and is stable inside the space $W$. The 
Cases refer to \S\ref{ss:geom} below. 
\begin{eqnarray*}
{\rm Case \ 1} \ \ \widehat B_2: \mathcal{F} &=& -q_3 + \varepsilon(u^2 \pm v^2 + \varepsilon +  q_1 ) \\
{\rm Case \ 2} \ \ \widehat B_3: \mathcal{F} &=& -q_3 + \varepsilon(u^2 \pm v^2 \pm \varepsilon^2 + q_2 \varepsilon  + q_1), \\ 
{\rm Case \ 3} \ \ \widehat C_3 : \mathcal{F} &=& -q_3 + \varepsilon(u^3 + u\varepsilon + \varepsilon + q_2 u + q_1 \pm v^2). 
\end{eqnarray*}
\end{prop}

When we consider 1-parameter families of surfaces we are permitted to impose one extra condition on the derivatives of the surfaces.  This gives rise to the following six additional cases: 

\begin{prop}\label{prop:family} {\rm (See Proposition~\ref{C3-2-proof})} \ \ 
For a generic one-parameter family of pairs of surfaces  $M_\tau,N_\tau$ near $(0,1,0)\in N_0$, in addition to the list of cases from Proposition~\ref{prop:2surfaces} the affine generating family germ $\Phi$ is space-time contact equivalent to one of the following normal forms and is stable inside the space $W$.
\begin{eqnarray*}
 {\rm Case \ 2a} \ \ \ \ \widehat B_4 : \mathcal{F} &=& -q_3 + \varepsilon(u^2 \pm v^2 + \varepsilon^3 + \tau\varepsilon^2 + q_2 \varepsilon + q_1), \\
{\rm Case \ 3a /  3e } \ \ \widehat C_3^{*} :  \mathcal{F} &=& -q_3 + \varepsilon(u^3 + u\varepsilon + \varepsilon + (\tau \pm q_2^2)u + q_1 \pm v^2), \\
 {\rm Case \ 3b} \ \ \ \ \widehat C_4:  \mathcal{F} &=& -q_3 + \varepsilon(u^4 + \tau u^2 + u\varepsilon + \varepsilon +  q_2 u + q_1 \pm v^2), \\
{\rm Case \ 3c} \ \ \widehat C_{3, 1} :  \mathcal{F} &=& -q_3 + \varepsilon(u^3 + u(\tau\varepsilon \pm \varepsilon^2) + \varepsilon +  q_2 u + q_1 \pm v^2), \\
{\rm Case \ 3d} \ \ \ \ \widehat F_4 :  \mathcal{F} &=& -q_3 + \varepsilon(u^3 + u\varepsilon \pm \varepsilon^2+ \tau\varepsilon +  q_2 u + q_1 \pm v^2), 
 \end{eqnarray*}
 and the non-simple 
 \begin{eqnarray*}
  {\rm Case \ 2b} \ \ \   \widehat B_3^{**} :  \mathcal{F} &=& 
    -q_3 + \varepsilon(u^2 \pm v^2 \pm  \varepsilon^2 +  a(q_1, q_2) \varepsilon  \pm q_1^2 \pm q_2^2 + \tau)    \
    \end{eqnarray*}
 where $a(q_1,q_2)$ is a functional modulus. 
 \end{prop}
 
See Figure~3 for an adjacency diagram of these singularities.

\begin{remark}\label{rem:normsforms}
{\rm
The notation used for these normal forms varies in the literature (compare \cite{GZ1} and \cite{RZ1}). 
Some of these here are new classes, but we opt to follow the naming convention used in \cite{RZ1}.  
The \ $\widehat{~} $ \ 
 indicates that the singularity occurs in the vicinity of the surface $N$ or $N_0$,  which is the case for all singularities considered in this article. For the unstarred
 singularities, the numerical subscripts refer to the codimension of the singularity. 
 In each case the signs of the $\pm$ are independent and correspond to different singularity types.  Occasionally we distinguish two distinct sub-cases of $\widehat{C}_3^*$  as $\widehat{C}_3^{*+}$ or $\widehat{C}_3^{*-}$ (see Proposition \ref{C3-2-proof}).  In all cases the sign of
$v^2$ gives different singularity types but does not affect the criminant up to local diffeomorphism.
The ${~}^*$ indicates that the singularity of the same name fails to be versally unfolded in the `standard' way by terms linear in $q_i$ and $\tau$. See Remark~\ref{rem:B3}
for a note about the apparently missing $\widehat{B}_3^*$
which fails to occur in our geometrical context.
}
\end{remark}

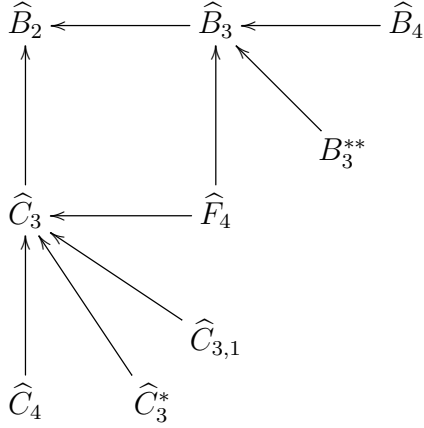
\begin{figure}\label{Adj_1}
\begin{center}
\ \ \ \ \ \ \ \ \ \ \ \ \ \ \ \ \ \xymatrix@!0{
\widehat{B}_2
& &  &   \widehat{B}_3 \ar[lll] & & &  \widehat{B}_4 \ar[lll] & & & 
\\
 \\  
& &  &  & &   B_3^{**}  \ar[uull]    \\
\widehat{C}_{3}  \ar[uuu]
& & &     \widehat{F}_{4} \ar[lll] \ar[uuu]  & &   
\\
\\
& &  & 
 \widehat{C}_{3,1} \ar[llluu] & & & &    & & &   \\  
 \widehat{C}_{4} \ar[uuu] & & \widehat{C}^{*}_{3} \ar[lluuu] } \\

\caption{\small An adjacency diagram of the various singularity types.} 
\end{center}
\end{figure}

\section{Geometrical properties}\label{ss:geom}
Here we consider the local geometry of $M$ at $(0,0,0)$ and $N$ at $(0,1,0)$, 
but the singularity that we refer to below is the one that occurs on $N$.  We relate
the different cases to the normal forms  listed in Propositions~\ref{prop:2surfaces}
and~\ref{prop:family}. The tie line corresponding to the base points on $M$
and $N$ is the second coordinate axis in $\RR^3$,
joining the points $(0,0,0)\in M$ and $(0,1,0)\in N$.

The calculations for generic surfaces $M,N$ (Proposition~\ref{prop:2surfaces}) are
performed using the direct method of \S\ref{ss:equations}.
For generic families $M_\tau, N_\tau$ (Proposition~\ref{prop:family}) the direct
method is used to analyse the geometry of the base surfaces $M_0, N_0$
but for the evolution of criminant and binodal
curves for $\tau\ne 0$ we need the method of normal forms;
see \S\ref{ss:representation}. The details of a sample reduction to normal form are given
in \S\ref{s:proofs}.

\mn
Case 1: $\widehat B_2$. The simplest case, where $M$ is not parabolic at $(0,0,0)$ and
$g_{02}\ne 0$, so that the tie line is not an asymptotic direction on $N$ at
$(0,1,0)$. Both binodal curves are smooth.

\mn
Case 2: $\widehat B_3$. Suppose $M$ is not parabolic at $(0,0,0)$ but  we
impose the single condition $g_{02}=0$. We require
$g_{03}\ne 0, g_{11}\ne 0$, that is
the tie line is asymptotic at $(0,1,0)\in N$ but $(0,1,0)$ is not a flecnodal point of $N$; 
also $N$ is not parabolic at $(0,1,0)$.
Both binodal curves, in $M$ and in $N$, are smooth and the tangent
to the binodal curve in $N$ is  along the tie line. (If an asymptotic direction
on $M$ is along the tie line then also the tangent to the binodal curve in $M$ is
along this line.)

\mn
Case 2a: $\widehat B_4$. Suppose $M_0$ is not parabolic at $(0,0,0)$, but $g_{02}=g_{03}=0, g_{04}\ne 0,$
that is $(0,1,0)$ is a flecnodal point of $N_0$ with asymptotic direction  along the
tie line, but this does not have 5-point contact with $N_0$. 
As this is two differential conditions it is
stable only in a generic 1-parameter family of surface pairs $M_\tau$ and $N_\tau$ , and for
genericity here we also require $g_{11}\ne 0$ (equivalently, $(0,1,0)$ is
not a parabolic point of $N_0$) and $g_{021}\ne 0$.
Both binodal curves, in $M_0$ and in $N_0$, are smooth
(hence this remains true for $M_\tau$ and $N_\tau$) and the tangent
to the binodal curve in $N_0$ is  along the tie line. (If an asymptotic direction
on $M_0$ is along the tie line then also the tangent to the binodal curve in $M_0$ is
along this line.) The  criminant in this case is a swallowtail surface but the
binodal curve is the intersection of the criminant with $N_0$ and this curve
is smooth as in Case 2.  The criminant, that is the ruled surface of
tie lines, is of course tangent to $N_0$ and $N_\tau$ along the binodal curve.

\mn
Case 2b: $\widehat B_3^{**}$. Suppose $M_0$ is not parabolic at $(0,0,0)$, but 
$g_{02}=g_{11}=0$, thus (0,1,0) is parabolic on $N_0$ and the asymptotic
direction there is along the tie line. 
We assume also $g_{20}\ne 0, g_{03}\ne 0$,
so that $(0,1,0)$ is not a flat umbilic on $N_0$ and also is not a flecnodal point,
which here is equivalent to saying that it is not a cusp of Gauss. Using the direct method of \S\ref{ss:equations} it
can be checked that
the binodal curve on $M_0$ is either an isolated point or two curves having
3-point contact (that is having the same 2-jet), while on $N_0$ there is respectively an isolated point or a
transverse crossing of two smooth curves. This method
does not allow us to determine the behaviour as $\tau$ passes
through 0 in a generic family. For the evolution of the
binodal curve on $M_\tau$ see Case 3e. Examples suggest that
the evolution on $N_\tau$ is a Morse transition: an isolated
point will disappear or evolve into a smooth curve, while a crossing
resolves into two smooth curves (e.g. by moving the asymptotic direction away
from the tie line, giving Case 2a). Compare~\cite[Prop.2.7]{BGT}.

\mn
Case 3: $\widehat C_3$. Suppose $M$ is parabolic at the origin,  but not a cusp of Gauss,
and the asymptotic direction at the origin is not along the tie line. 
Thus writing the quadratic terms of $M$ as $(au+v)^2$, the linear
form $au+v$ is not a factor of the cubic terms of $M$.
Suppose
also that $g_{02}\ne 0$, so that the tie line is not an asymptotic direction for $N$,
that $(0,1,0)$ is not a parabolic point on $N$ and that $f_{11}g_{02}\ne f_{02}g_{11}$. Then the binodal curve on $M$ is smooth, that on $N$ has an ordinary cusp
(locally diffeomorphic to $\{(t^2,t^3)\}$)
with the limiting tangent at the cusp not along the asymptotic direction at $(0,1,0)$.
 
\mn
Case 3a:  $\widehat C_3^{*}$. Suppose $M_0$ is parabolic at $(0,0,0)$, $N_0$ is parabolic at $(0,1,0)$, neither has
a cusp of Gauss, neither has the asymptotic direction along the tie line (that is
$f_{02}$ and $g_{02}$ are both nonzero), and these asymptotic directions are
not parallel (in the presence of the two previous conditions
this is equivalent to $f_{11}g_{02}-f_{02}g_{11}\ne 0$). This is a symmetrical
situation between $M_0$ and $N_0$, and occurs generically in a 1-parameter family.  Here we distinguish two sub-cases as $\widehat C_3^{*+}$ and $\widehat C_3^{*-}$  (see Proposition \ref{C3-1-proof}). 
On $M_0$ and $N_0$ the binodal curve consists of two branches having
exactly 3-point contact ($\widehat C_3^{*-}$), or else is an isolated point  ($\widehat C_3^{*+}$). Writing the quadratic terms
for $M_0$ as $(au+v)^2$ and those for $N_0$ as $(cx+y)^2$ the condition for
two real branches is 
\begin{equation}\label{eq:beaks}
(f_{30}-f_{21}a+f_{12}a^2-f_{03}a^3)(g_{30}-g_{21}c+g_{12}c^2-g_{03}c^3)>0.
\end{equation}
In a generic family two real branches separate as a `beaks' transition and an isolated
point evolves as a `lips' transition. See \S\ref{s:proofs}. The criterion 
(\ref{eq:beaks}) above for a
two real branches coincides with the condition $c_3<0$ for a
beaks transition in \S\ref{s:proofs}.  The criminant is locally diffeomorphic to a folded Whitney umbrella.  

\mn
Case 3b: $\widehat C_4$. Suppose $M_0$ has a 
(nondegenerate) cusp of Gauss at $(0,0,0)$, $N_0$ is not parabolic
at $(0,1,0)$,  $g_{02}\ne 0$ (that is the tie line is not in an asymptotic
direction) and $f_{11}g_{02}-f_{02}g_{11}\ne 0$. 
The binodal curve in $M_0$ is then smooth, with tangent the asymptotic direction
which is also the tangent to the parabolic curve on $M_0$.
The binodal curve in $N_0$ has a cusp locally
diffeomorphic to $(t^3,t^4)$, In a generic family this evolves as a
`swallowtail transition'.  The extra feature in this case is a `local binodal curve'
on $M_0$: there are bitangent planes of $M_0$ whose contact points tend to coincidence on $M_0$, forming a smooth curve
passing through the cusp of Gauss
and tangent to the parabolic curve there. This `local binodal curve' lies in the
hyperbolic region of $M_0$, but the bilocal binodal curve defined by bitangent
planes having contact points one on $M_0$ and 
the other on $N_0$  need not. 

The relative positions
of the local binodal curve, the bilocal binodal curve and the parabolic curve
depend on the local geometry of $M_0$. In fact taking the surface $M_0$ to have
local form $z=y^2 + f_{21}x^2y+f_{12}xy^2+f_{03}y^3+...+f_{40}x^4+\ldots$
the leading terms in the expressions for these three curves on $M_0$ are:
\[ \mbox{Local: } y=-\frac{2f_{40}}{f_{21}}x^2, \ \mbox{ Bilocal }:
y=-\frac{f_{21}}{2}x^2, \ \mbox{ Parabolic } y=-\frac{6f_{40}}{f_{21}}x^2.   \]
A nondegenerate cusp of Gauss in these coordinates has $f_{21}^2\ne 4f_{40}$.
If $f_{40}<0$ the cusp of Gauss is automatically elliptic ($f_{21}^2>4f_{40}$),
the bilocal binodal curve is in the hyperbolic region of $M_0$ (locally $f_{21}x+\mbox{ h.o.t. in }
x,y \mbox{ is } <0$ for hyperbolic points of $M_0$) and the local curve separates it from
the parabolic curve. The situation if $f_{40}>0$ is more complicated.

\mn
Case 3c:  $\widehat C_{3, 1}$. Suppose $(0,0,0)$ is a parabolic point of $M_0$ 
(not a cusp of Gauss), $(0,1,0)$ is not a parabolic point of $N_0$,
the tie line is not in an asymptotic direction on $N_0$ (that is $g_{02}\ne 0$)
 and  $f_{11}g_{02}=f_{02}g_{11}$.  Then on $N_0$ the binodal curve has an ordinary cusp with the limiting tangent to the cusp parallel to
the asymptotic direction on $M_0$.  The binodal curve on $M_0$ is smooth with
tangent also parallel to the asymptotic direction.  

\mn 
Case 3d:  $\widehat F_4$.
Suppose $(0,0,0)$ is a parabolic point of $M_0$ 
(not a cusp of Gauss), $(0,1,0)$ is not a parabolic point of $N_0$,
the tie line is in an asymptotic direction on $N_0$ but not at a
flecnodal point, that is $g_{02}=0, g_{03}\ne 0$. Finally suppose 
$f_{11}g_{02}\ne f_{02}g_{11}$, that is  $f_{02}g_{11}\ne 0$.
Geometrically this is the same as (3c):  on $N_0$ the binodal curve has an ordinary cusp with the limiting tangent to the cusp parallel to
the asymptotic direction on $M_0$.  The binodal curve on $M_0$ is smooth with
tangent also parallel to the asymptotic direction. The criminant is locally diffeomorphic to an open swallowtail.  

\mn
Case 3e:  $\widehat C_3^{*}$. This is the same as Case 2b, with
$M_\tau$ and $N_\tau$ reversed, and the singularity
referring as always to the point $(0,1,0)\in N_0$. Thus
$N_0$ is not parabolic at $(0,1,0)$, $f_{02}=f_{11}=0$, while
$f_{20}$ and $f_{03}$ are nonzero. On $N_0$ the binodal
curve is an isolated point or two curves with 3-point contact
and the evolution of the binodal curve as $\tau$ 
passes through zero is via respectively
a lips or beaks transition.

\subsection{Two proofs}\label{ss:geomproofs}
There follow two sample proofs of the geometrical statements
above.

\medskip\noindent
{\bf Case 3a} Since we are looking at the single surface $N_0$
we can use the forms (\ref{eq:local}) and because of the
assumptions in this case we can write
\begin{eqnarray*}
f(u,v) &=& (au+v)^2 + f_{30}u^3+f_{21}u^2v+f_{12}uv^2+f_{03}v^3+ \ldots \\
g(x,y)&=&(cx+y)^2+g_{30}x^2+g_{21}x^2y+g_{12}xy^2+g_{03}y^3+\ldots
\end{eqnarray*}
 where $a\ne c$ since the asymptotic directions on $M_0$
 and $N_0$ are not parallel.  We shall also assume
 $a\ne 0$ and $c\ne 0$: these assumptions are not 
 necessary to the argument and in any case are generic
 assumptions.

 The conditions for the tangent planes at
 $(u,v,f(u,v))\in M_0$ and $(x,y+1,g(x,y))\in N_0$ to be identical are (subscripts denoting partial derivatives)
 \begin{eqnarray}\label{eq:3equations}
    {\rm (i)} \ \ \ \  f_u(u,v,f(u,v))&=&
    g_x({x,y,g(x,y)}) \nonumber \\
    {\rm (ii)} \ \ \  f_v(u,v,f(u,v))&=&
     g_y({x,y,g(x,y))})\\
    {\rm (iii)} \ \ \ (x-u)f_u(u,v,f(u,v))&+&(y+1-v)f_v(u,v,f(u,v))
    \nonumber \\
   &+& f_u(u,v,f(u,v))-g(x,y,g(x,y))=0 \nonumber
 \end{eqnarray}

The first two equations state that the normals to
$M_0$ and $N_0$ are parallel, so that the tangent
planes are parallel.
The third equation states that the line joining the
two points on $M_0$ and $N_0$ is perpendicular to
the normal to $M_0$, and therefore also to the
normal to $N_0$. The two tangent planes are therefore
identical.

Solving (i) for $u$ as a function of $v,x,y$ and substituting
in (ii) and (iii) gives equations with linear terms
\[ \frac{2c(c-a)}{a}x +\frac{2(c-a)}{a}y \ \ 
\mbox{ and } \ \ \frac{2c^2}{a}x+\frac{2c}{a}y\]
respectively. Because these are proportional we cannot use
them to solve for $x$ and $y$ as functions of $v$ but instead
solve (ii) for $x$ as a function $x(v,y)$ and substitute
further in (iii). The result is an expression of the form
\[ x = -\frac{y}{c} + \alpha v^2 + \beta y^2 + \ldots\]
for certain values of $\alpha$ and $\beta$ and
an equation in $v$ and $y$ with
2-jet
\[\frac{3(a^3f_{03} - a^2f_{12} + af_{21} - f_{30})}{a^2(a - c)}v^2 - \frac{3(c^3g_{03} - c^2g_{12} + cg_{21} - g_{30})}{c^2(a -c)}y^2.\]
Neither coefficient is zero since neither surface has a cusp of
Gauss: the condition for this is that respectively
$(au+v)$ and $(cx+y)$ should be a factor of the cubic
terms of $f$ and $g$. If the two coefficients of $v^2$ and $y^2$
have opposite sign then the locus in $v,y$ has an isolated point,
so the same applies to the locus in $x,y$ which we are studying
here, so we assume the signs are the same and write
the 2-jet as $A^2v^2-B^2y^2$ where $A>0, B>0$. The two branches
in the $v,y$ plane therefore have the form say
\[ v_1=\frac{B}{A}y + v_{12}y^2+v_{13}y^3+\ldots \mbox{ and }
v_2=-\frac{B}{A}y + v_{22}y^2+v_{23}y^3+\ldots.\]
These can then be substituted into $x(v,y)$ as above to
obtain $x$ as a function of $y$ alone for the two
branches of the binodal curve in the $x,y$ plane. To show
that the branches in the $x,y$ plane have at least 3-point
contact we do not need to calculate further for substituting
for $v=v_1$ and $v=v_2$ yields 
\[x=-\frac{y}{c}+\alpha\left(\frac{B}{A}y +v_{12}y^2\right)^2 +
\beta y^2 +\ldots \mbox{ and } x=-\frac{y}{c}+\alpha\left(-\frac{B}{A}y +v_{22}y^2\right)^2 +
\beta y^2+\ldots\]
respectively, which clearly have the same 2-jet.

It takes a little more effort to show that the contact is
exactly 3-point but calculating $v_{12}$ and $v_{22}$ and remembering
the definitions of $A$ and $B$  the difference between the
coefficients of $y^3$ in the branches of the binodal
curve in the $x,y$ plane comes to
 a nonzero multiple of
 \[(a-c)(a^3f_{03} - a^2f_{12} + af_{21} - f_{30})^2(c^3g_{03} - c^2g_{12} + cg_{21} - g_{30})\]
 which is known to be nonzero.\hfill$\Box$

 \bigskip
 \noindent
 {\bf Cases 2b and 3e} We shall use the notation of Case 2b but 
 the singularity on $N_0$ for Case 3e is the same as that
 on $M_0$ for Case 3e and the calculation will find both. The
 calculation below is less detailed than Case 3a above.

 From the conditions of Case 2b the  functions $f$ and $g$
 as in (\ref{eq:local}) take the form
 \begin{eqnarray*}
    f(u,v) &=& f_{20}u^2+f_{11}uv+f_{02}y^2 + f_{30}u^3+f_{21}u^2v+f_{12}uv^2+f_{03}v^3+ \ldots \\
g(x,y)&=&g_{20}x^2+g_{30}x^2+g_{21}x^2y+g_{12}xy^2+g_{03}y^3+\ldots
 \end{eqnarray*}
where $f_{11}^2-4f_{20}f_{02}, \ \  g_{20}$ and $g_{03}$ are all nonzero. We shall also assume for the calculation that $f_{11}\ne 0$: this condition can be avoided by taking a different route.

The three equations (\ref{eq:3equations}) allow us to first express
$u$ as $u(v,x,y)$ using (i), then $x$ as $x(v,y)$ using (ii); we end up with a relationship between $v$ and $y$ without linear terms,
of the form
\begin{eqnarray}\label{eq:pqr}
pv^2+qvy+ry^2+ \mbox{ higher terms,  where }  
 q=\frac{g_{12}(f_{11}^2-4f_{20}f_{02})}{f_{11}g_{20}}, \ \ r=3g_{03}.
\end{eqnarray}
Here $p$ is a much more complicated expression in the second
and third order terms of $f$ and $g$; however $p$ is not in fact
needed for the calculation here.

We assume that (\ref{eq:pqr}) has two distinct real branches, that is
$p^2-4qr> 0$, otherwise the
corresponding binodal curves in the $(u,v)$ and $(x,y)$ planes
will be isolated points. The case of tangential branches is ruled
out as non-generic. Write
\begin{eqnarray}\label{eq:alphabeta}
   y=y_i(v) = \alpha_i v + \beta_i v^2 + \ldots, \ \ i=1,2 
\end{eqnarray} 
for these two branches, where $\alpha_1\ne
\alpha_2$. We want to deduce, from the locus (\ref{eq:pqr}) in the $(v,y)$ plane, the corresponding loci in the $(u,v)$ plane
(for $M_0$) and in the $(x,y)$ plane (for $N_0$).

For $M_0$, that is in the $(u,v)$ plane, the expression for 
$u(v,y)$ has the form
\begin{eqnarray}\label{eq:abc}
-\frac{2f_{02}}{f_{11}}v+av^2+bvy+cy^2+ \mbox{ higher terms,  where } 
b=\frac{g_{12}(f_{11}^2-4f_{20}f_{02})}{f_{11}^2g_{20}}, \ \ 
c=\frac{3g_{03}}{f_{11}}
\end{eqnarray}
and $a$ is more complicated but does not in fact enter the
calculation. Thus the coefficients of $v$ in the expressions
for $u(v,y_1(v))$ and $u(v,y_2(v)) $ will be equal, so the branches
in the $(u,v)$ plane will be tangent. The coefficients
of $v^2$ in these two branches are
$a+b\alpha_1+c\alpha_1^2$ and $a+b\alpha_2+c\alpha_2^2$.
These are equal if and only if $(\alpha_1-\alpha_2)(b+c(\alpha_1+\alpha_2))=0.$  But $\alpha_1$ and $\alpha_2$ satisfy the equation
$p+q\alpha+r\alpha^2=0$, by (\ref{eq:pqr}), so $\alpha_1+\alpha_2=-q/r$,
and equality of the 2-jets requires $qc=br$, which is easily checked
to be true.  Hence the branches in the $(u,v)$ plane have
(at least) 3-point contact. The condition for exactly 3-point
contact is too complicated to reproduce here. This completed the proof
for $M_0$.

\medskip

For the binodal curve on $N_0$ let us write the two branches in the
$(v,y)$ plane, assumed real, as in (\ref{eq:alphabeta})
above. Now $x$ as a function of $v$ and $y$ has linear term $Av$ where
$A=4(f_{11}^2-4f_{20}f_{02})/2f_{11}g_{20}$. Thus the linear terms
of the $(x,y)$ curve, parametrised by $v$, are $(Av,\alpha_1v)$
and $(Av,\alpha_2v)$ The branches  therefore form a transverse
crossing since $\alpha_1\ne\alpha_2$.  This completes the calculation
for $N_0$ and therefore for Cases 2b and 3e. \hfill$\Box$
\section{Representation of the criminant surface and the binodal
curves using normal forms for the generating functions $\FF$}
\label{ss:representation}

In \S\ref{s:proofs} we present explicit calculations and derive the normal forms for Case 3 and Case 3a (see 
 Propositions~\ref{prop:2surfaces} and ~\ref{prop:family}, and \S\ref{ss:geom}).
 In the present section we use the normal
 forms to draw the criminant surface and
 the binodal curves.
 
The reason for selecting these cases is that
they  are referred
to as K2 and K3 by Korteweg in
 \cite{K2}; see also `Second theorem' and `Third theorem' in \cite[p.76-77]{Sengers}. The `second theorem' (K2) has already
 been noted in Remark~\ref{rem:K2}.
 The `third theorem' (K3) asserts that
  for our Case 3a, at the moment of
  transition $\tau=0$ in the family of surfaces $M_\tau, N_\tau$ ,
  the binodal curves on both surfaces
  $M$ and $N$ consist of two `real or imaginary branches' having the same tangent
  and curvature. The case of real branches
  we refer to as a `beaks transition' 
  ($\widehat C_3^{*-}$); imaginary branches give a `lips
  transition', 
  $\widehat C_3^{*+}$, where the real binodal curves
  are isolated points. As the family of
  surfaces evolves, the `lips' becomes
  empty in one direction and in the other
  direction opens out in the manner shown
  in Figure~\ref{C3*plus_crim}. However
  this figure suggests that the `lips'
  curve has 4 inflexions. As noted in
  Remark~\ref{rem:geom}, our methods do not preserve inflexions, but according to Korteweg's own calculations there are actually two inflexions, as in Figure~\ref{K-lips}.

First we show how the normal form can be used to derive an explicit local parametrisation for the criminant  surface, that is the ruled surface formed by the tie ines.  We do the calculations for the Case 3 $(\widehat C_3$) but the other parametrizations can be derived similarly.  We also include figures for the two Cases 3 and 3a.

The criminant $\Delta\mathcal{F}$ for a normal form  $\mathcal{F}$ as in Proposition~\ref{prop:2surfaces} is the set given by  \[ \Delta\mathcal{F} = \bigg\{ (q_1, q_2, q_3) \ | \  \exists (u, v, \varepsilon), \mathcal{F}=\frac{\partial \mathcal{F}}{\partial u}=\frac{\partial \mathcal{F}}{\partial v}=\frac{\partial \mathcal{F}}{\partial \varepsilon}=0 \bigg\}. \]
The problem here is to turn this description as a zero set into a {\em local parametrization} close to the base point on $N$.

For the normal form 
\[\widehat C_3 : \mathcal{F} = -q_3 + \varepsilon(u^3 + u\varepsilon + \varepsilon + q_2 u + q_1 \pm v^2)\]
we calculate the derivatives as
\begin{eqnarray*}
\frac{\partial \mathcal{F}}{\partial u} &=& \varepsilon(3u^2 + \varepsilon + q_2) \\
\frac{\partial \mathcal{F}}{\partial v} &=& \pm 2 \varepsilon v, \\
\frac{\partial \mathcal{F}}{\partial \varepsilon} &=& u^3 + 2\varepsilon u +2\varepsilon + q_2 u + q_1 \pm v^2.
\end{eqnarray*}

 The vanishing of $\frac{\partial \mathcal{F}}{\partial v}=\pm 2 \varepsilon v$ means that either  $\varepsilon=0$ or $v=0$ if $\varepsilon \neq 0$.
 
 \begin{enumerate}
     \item  If $\varepsilon=0$, then $\frac{\partial \mathcal{F}}{\partial u} $ is automatically zero and $\mathcal{F} = 0$ implies $q_3=0$.  Then $\frac{\partial \mathcal{F}}{\partial \varepsilon}=0$ gives that $q_1 = -u^3 - 2\varepsilon u -2\varepsilon - q_2 u  \mp v^2$ where $q_2$ and the other variables are arbitrary. So this component, called the redundant component, is given by the plane $q_3=0$ and it corresponds to the surface $N$.  
     
     \item If $\varepsilon \neq 0$ then $\frac{\partial \mathcal{F}}{\partial u}=0$ gives that $q_2= - 3u^2 - \varepsilon$,  and $\frac{\partial \mathcal{F}}{\partial \varepsilon}$  gives $q_1=2u^3 - \varepsilon u  - 2 \varepsilon$. Substituting these into $\mathcal{F} =0$ gives $q_3=-\varepsilon^2 u - \varepsilon^2$, so we have ${(q_1, q_2, q_3) = 2 u^3 - 2\varepsilon, -3 u^2,  -\varepsilon^2}$ which is a folded Whitney umbrella.
 \end{enumerate}

The curve along which the folded Whitney umbrella of the criminant and the surface $N$ are tangent is given by $(0,2u^3,-3 u^2)$. This corresponds to the binodal curve and has an ordinary cusp at the origin.

\begin{figure}[H]
\centering
\includegraphics[width=1\textwidth]{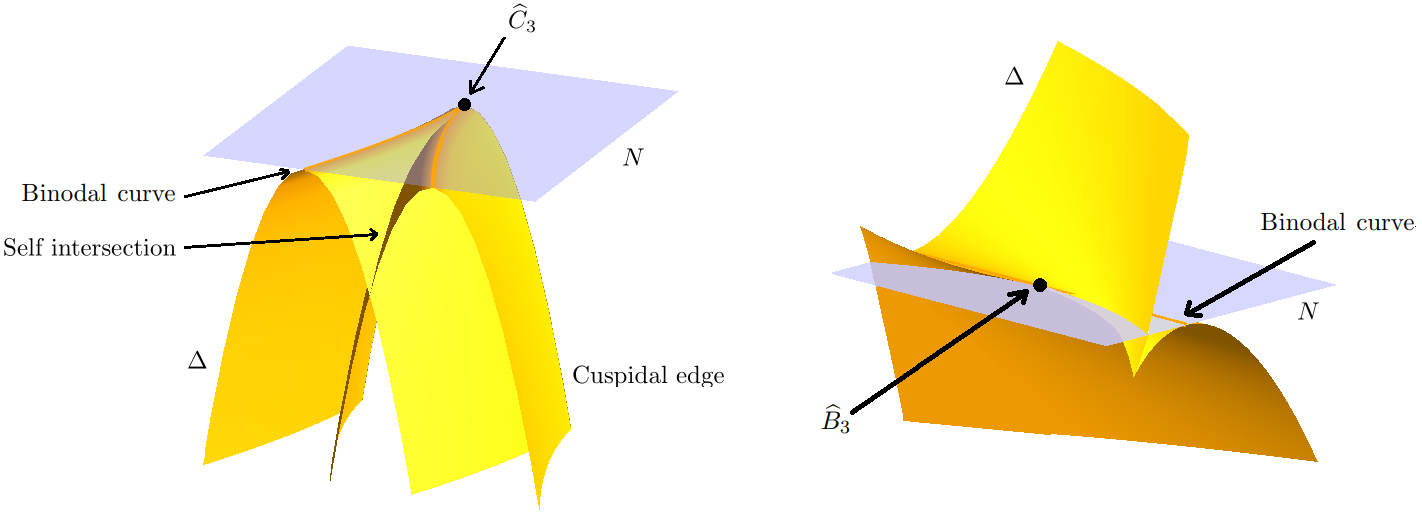}
\caption{\small The criminants and binodal curves of $\widehat C_3$ and $\widehat B_3$, drawn from the normal forms as in Proposition~\ref{prop:2surfaces}.
In the left-hand figure the curved surface
is called a `folded Whitney umbrella'.
Note that the surface $N$ becomes a plane in this representation, so geometric information about
the curvature of $N$ is lost. See
Remark~\ref{rem:geom} for information about what is lost
and what is preserved by using normal forms.}
\label{C3_and_B3_crim}
\end{figure}

\begin{figure}[H]
\centering
\includegraphics[width=1\textwidth]{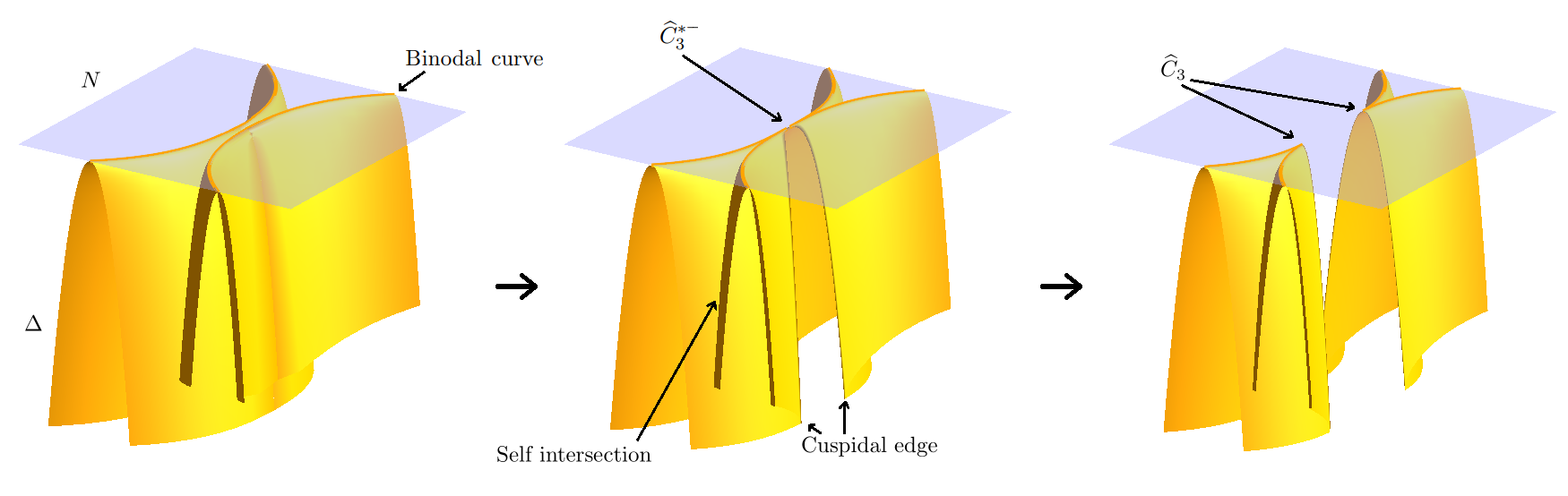}
\caption{\small The criminants and binodal curves  of $\widehat C_3^{*-}$
as it evolves (left to right) in a generic 1-paramter family of surfaces,
using the normal form as in 
Proposition~\ref{prop:family}.}
\label{C3*minus_crim}
\end{figure}

\begin{figure}[H]
\centering
\includegraphics[width=1\textwidth]{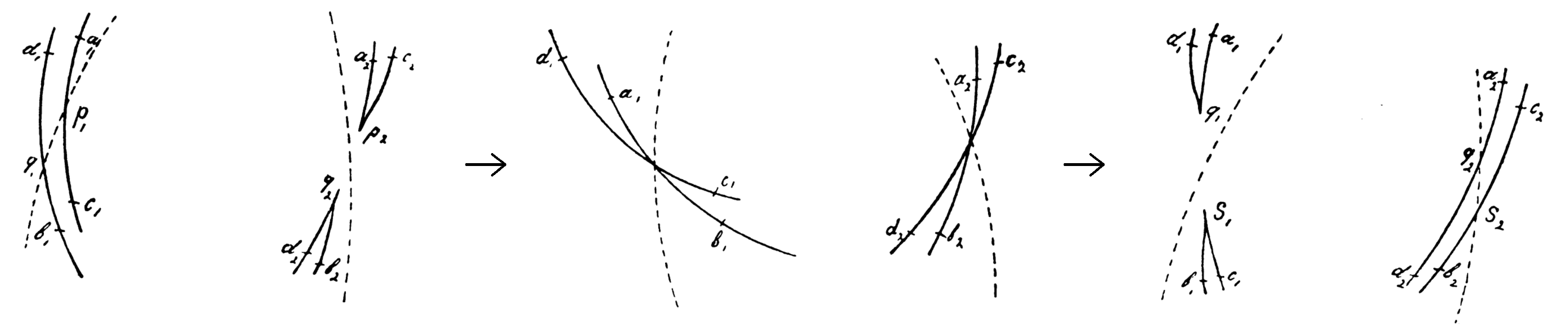}
\caption{\small The binodal curves during a beaks transition of Case 3a ($\widehat C_3^{*-}$) from Korteweg 1891 \cite{K4}, also reproduced in \cite[p.79]{Sengers}.  The transitions taking place at both $M$ and $N$ are shown simultaneously. The centre pair in this diagram
show two curves (full black lines) which should be
tangential as well as crossing: they have 3-point
contact in the manner of $y=\pm x^3$.
The dashed lines represent the parabolic (spinodal) curve.}
\label{K-beaks}
\end{figure}

\begin{figure}[H]
\centering
\includegraphics[width=0.95\textwidth]{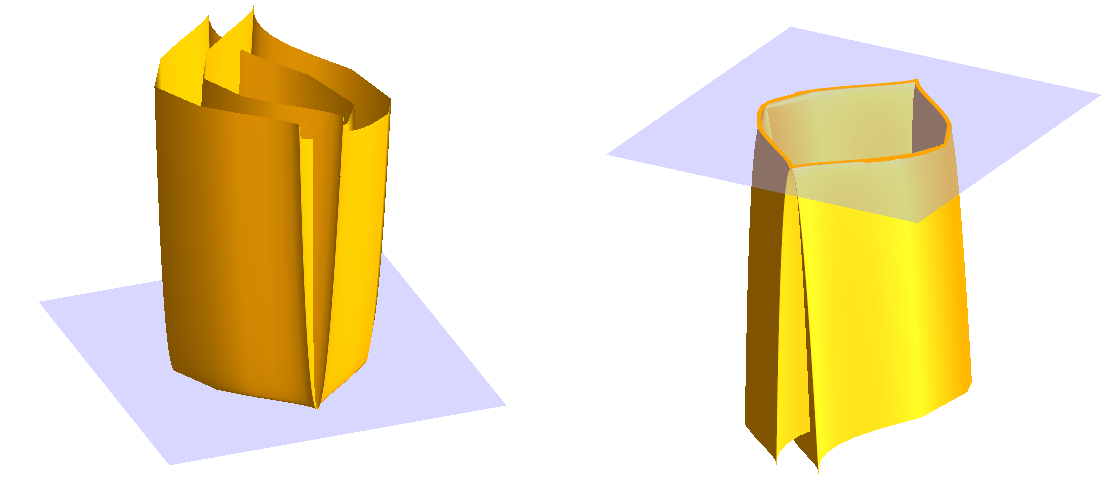}
\caption{\small Two views of the criminant of $\widehat C_3^{*+}$ for a small negative value of $\tau$. The `lips' figure can be
seen on the right, but beware that 
inflexions are not preserved by our
methods so the presence of 4 inflexions
rather than 2 as in the next figure
is not significant.}
\label{C3*plus_crim}
\end{figure}

\begin{figure}[H]
\centering
\includegraphics[width=0.4\textwidth]{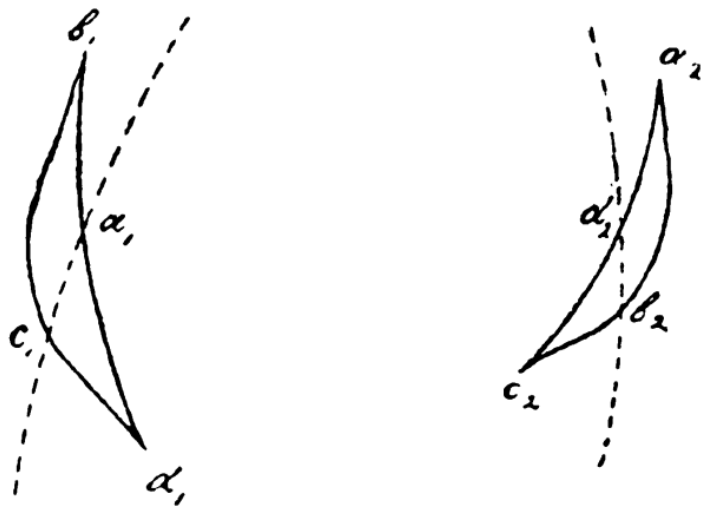}
\caption{\small A lips transition of the binodal curve in Case 3a (for a small negative value of $\tau$ in $\widehat C_3^{*+}$) from  Korteweg 1891 \cite{K4}. (This figure
is not reproduced in \cite{Sengers}.) The binodal curves at both $M$ and $N$ are shown simultaneously. Note
that our methods in this article cannot verify the 
presence of {\em inflexions} in this figure. Compare
Remark~\ref{rem:geom}.}
\label{K-lips}
\end{figure}

 \section{Reduction to normal forms, calculations and proofs}\label{s:proofs}

For Case 3 we have the following proposition:

\begin{prop}
For a pair of surfaces $M$ and $N$ corresponding to Case 3, 
near $\lambda = 0$ the affine generating family germ $\Phi$ is space-time contact equivalent to the following normal form and is stable inside the space $W$:  \[\widehat C_3 : \mathcal{F} = -q_3 + \varepsilon(u^3 + u\varepsilon + \varepsilon + q_2 u + q_1 \pm v^2).\]
\label{C3-1-proof}
 \end{prop}

 \begin{cor}
If $M$ has a parabolic point, the corresponding binodal curve on $N$ has an ordinary cusp. 
 \label{Cor_Lips}
 \end{cor}

For Case 3a, where both surfaces have parabolic points, the singularity $\widehat C_3$ fails to be versally unfolded and results in a singularity of type $C_3^{*+}$ or $C_3^{*-}$ depending on the sign of the expression  
\[c_3:=-\left(f_{30} - \frac{f_{21}f_{11}}{2f_{02}} + \frac{f_{12}f_{11}^2}{4f_{02}^2}  -\frac{f_{03}f_{11}^3}{8f_{02}^3}\right)\left(g_{30} - \frac{g_{21}g_{11}}{2g_{02}} + \frac{g_{12}g_{11}^2}{4g_{02}^2}  -\frac{g_{03}g_{11}^3}{8g_{02}^3} \right).\]

In particular we have the following proposition:
\begin{prop}
For a generic one-parameter family of pairs of surfaces $M_\tau$ and $N_\tau$ in which at the moment $\tau=0$ in the family, there exists a bitangent plane which is tangent to each surface at a parabolic point, are both parabolic, and share a    
near $\lambda = 0$ the affine generating family germ $\Phi$ is space-time contact equivalent to one of the following normal forms and is stable inside the space $W$:  \[\widehat C_3^{*+} : \mathcal{F} = -q_3 + \varepsilon(u^3 + u\varepsilon + \varepsilon + (\tau + q_2^2)u + q_1 \pm v^2), \  {\rm if} \ c_3 >0\]
\[\widehat C_3^{*-} : \mathcal{F} = -q_3 + \varepsilon(u^3 + u\varepsilon + \varepsilon + (\tau - q_2^2)u + q_1 \pm v^2), \  {\rm if} \ c_3 <0.\]
\label{C3-2-proof}
 \end{prop}
 
 The previous proposition together with explicit calculations from the normal form will prove:

 \begin{cor}
 The binodal curves on both $M$ and $N$ undergo a lips transition if $c_3>0$ and a beaks transition if $c_3 <0$. (See {\rm (\ref{eq:beaks})} in \S\ref{ss:geom} for a simplified version of $c_3<0$ when
 special coordinates are used.)
 \label{Cor_Lips}
 \end{cor}
  
  \noindent
{\bf Proof of Proposition \ref{C3-1-proof}.}
  In this case since $M$ is parabolic, we can make a change of coordinates $v=v' -  \frac{ u f_{11}}{2f_{02}}$ to reduce the part of the generating family that is quadratic in $u$ and $v$ and linear in $\varepsilon$ to give 
\begin{eqnarray*}
\Phi_0(u, v', \varepsilon) &&=-q_3 + \varepsilon\left( f_{02} v'^2 + \frac{8 f_{02}^3 f_{30} - 4f_{02}^2f_{11} f_{21} + 2f_{02}f_{11}^2 f_{12} - f_{03} f_{11}^3}{8f_{02}^3} u^3 + ...\right) \\ && + \left(g_{02}v'^2 + \frac{f_{02}g_{11} - f_{11}g_{02}}{f_{02}}uv' + \frac{4f_{02}^2g_{20} - 2f_{02}f_{11}g_{11} + f_{11}^2g_{02}}{4f_{02}^2} u^2 - 2 g_{02} v'\right. \\&& \left. -\frac{f_{02}g_{11} - f_{11}g_{02}}{f_{02}} u + g_{02})\right) \varepsilon^2 +...   +  (g_{02}+g_{03} + ... )\varepsilon^3 +...
\end{eqnarray*}

 Using the lowest degree terms in $u$ and $v'$ that are linear in $\varepsilon$, 
 a further change in coordinates (and dropping
 the prime $'$) can then remove any other terms divisible by $u^2\varepsilon$ or $v'\varepsilon$  to give the pre-normal form
 \begin{eqnarray*}
\Phi_0(u, v, \varepsilon) &&=-q_3 + \varepsilon\left( f_{02} v^2 + \frac{8 f_{02}^3 f_{30} - 4f_{02}^2f_{11} f_{21} + 2f_{02}f_{11}^2 f_{12} - f_{03} f_{11}^3}{8f_{02}^3} u^3 + ...\right) \\ && + \left(g_{02} -\frac{f_{02}g_{11} - f_{11}g_{02}}{f_{02}} u\right) \varepsilon^2   + ... 
\end{eqnarray*}  
In Case 3 we assume that each of the coefficients here is nonzero, and the vanishing of the various individual expressions  
$8 f_{02}^3 f_{30} - 4f_{02}^2f_{11} f_{21} + 2f_{02}f_{11}^2 f_{12} - f_{03} f_{11}^3$, $f_{02}g_{11} - f_{11}g_{02}, g_{02}$, and $f_{02}$
gives the various sub-cases 3b, 3c, 3d and 3e respectively.

The family is infinitesimally space-time stable inside the space of functions $W$ if any germ divisible by $\varepsilon$ lies in the tangent space to the orbit restricted to $q = 0$
  \[T_{\widehat{C_3}}\mathcal{F} = \mathcal{O}_{u, v,\varepsilon} \bigg\{\mathcal{F}, \frac{\partial \mathcal{F}}{\partial u}, \frac{\partial \mathcal{F}}{\partial v}\bigg\} + \mathcal{O}_\varepsilon \bigg\{ \frac{\partial \mathcal{F}}{\partial \varepsilon} \bigg\}.\]
 
 We can remove the monomials corresponding to the shaded region of Figure \ref{Newton_C3}  up to space-time contact equivalence as follows.
 We have mod $T_{\widehat{C_3}^*}\mathcal{F}$: 
 
\begin{figure}
\centering
\includegraphics[width=0.8\textwidth]{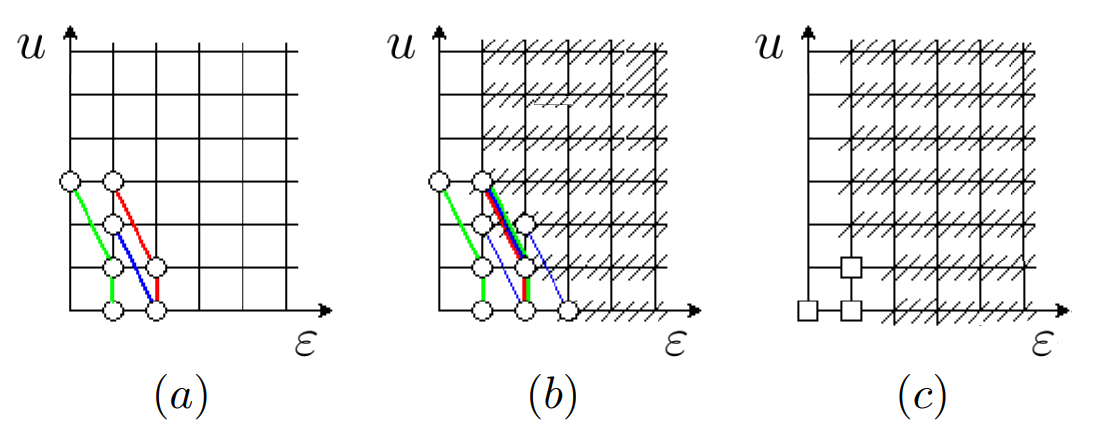}
\caption{\small The Newton diagram of $\widehat{C_3}$. }
\label{Newton_C3}
\end{figure}

 \begin{eqnarray}
 \mathcal{F} &=& \varepsilon(u^3 + \varepsilon u + \varepsilon + v^2) \equiv 0, \label{(9.1)} \\ 
  \frac{\partial \mathcal{F}}{\partial u} &=& 3u^2\varepsilon + \varepsilon^2  \equiv 0, \label{(9.2)}  \\
  \frac{\partial \mathcal{F}}{\partial v} &=& 2v\varepsilon  \equiv 0,  \label{(9.25)}  \\
   \frac{\partial \mathcal{F}}{\partial \varepsilon} &=& u^3 + 2 \varepsilon u + 2\varepsilon + v^2 \equiv 0. 
  \label{(9.3)}
   \end{eqnarray}

  We can multiply relation (\ref{(9.3)}) by $-\frac{1}{2} \varepsilon$ to give 
  
  \begin{equation}
  -\frac{1}{2} \varepsilon \frac{\partial \mathcal{F}}{\partial \varepsilon}  = -\frac{1}{2}  u^3 \varepsilon- \varepsilon^2 u- \varepsilon^2  -\frac{1}{2} v^2 \varepsilon \equiv 0 . \ \ \ \ \label{(9.4)}    
  \end{equation}

Firstly the relation (\ref{(9.25)}) implies that monomials divisible by $v\varepsilon$ belong to the tangent space to the orbit. The relations (\ref{(9.4)}) and (\ref{(9.1)}) then imply that monomials of the form $u^3 \varepsilon^{k_1}$ for some $k_1 \geq 0$ belong to the tangent space to the orbit. Using this fact, we can multiply (\ref{(9.2)}) by $u$ to show that monomials of the form $u \varepsilon^{k_2+2}$ for some $k_2 \geq 0$ also belong to the orbit. Now, multiplying relation (\ref{(9.3)}) by functions in $\varepsilon$ only, yield that $\varepsilon^{k_3+2}$ for some $k_3 \geq 0$. Now relation (\ref{(9.2)}) implies that monomials of the form $u^2 \varepsilon^{k_4}$ for some $k_4 \geq 0$ also belong. Finally, (\ref{(9.2)}) can be used to obtain any monomial in the shaded region of figure \ref{Newton_C3}. A basis for $Q = \varepsilon \mathbb{R}[[u, v, \varepsilon]]/T_{C_3}^*\Phi_0$ is generated by the three monomials $1, \varepsilon$ and $\varepsilon u$.

 Up to quadratic terms in $q$ the function $\Phi$ can now be written as
 \begin{equation}\label{eq:linearterms}
\Phi(u,v, \varepsilon) = \Phi_0(u,v, \varepsilon) + \phi_1(u,v, \varepsilon)q_1 + \phi_2(u,v, \varepsilon)q_2 + \phi_3(u,v, \varepsilon)q_3
\end{equation}
for some functions $\phi_1, \phi_2$ and $\phi_3 = -1$.

The singularity  $\widehat C_3$  is versally unfolded if the following matrix evaluated at the origin
\begin{equation}\label{eq:versalmatrix}
\begin{pmatrix}
\phi_1 & \phi_2 & \phi_3 \\
\frac{\partial \phi_1}{\partial \varepsilon} & \frac{\partial \phi_2}{\partial \varepsilon} & \frac{\partial \phi_3}{\partial \varepsilon}  \\
\frac{\partial^2 \phi_1}{\partial \varepsilon \partial u} & \frac{\partial ^2\phi_2}{\partial \varepsilon \partial u} & \frac{\partial^2 \phi_3}{\partial \varepsilon \partial u}
\end{pmatrix} =
\begin{pmatrix}
0 & 0 & -1 \\
g_{11} & 2g_{02}& 0   \\
-2g_{20} & -g_{11} & 0
\end{pmatrix}.
\end{equation}
has nonzero determinant.  So if $N$ is not a parabolic point, then  the singularity is versally unfolded and has normal form:

\[\widehat C_3 : F = -q_3 + \varepsilon(u^3 + u\varepsilon + \varepsilon + q_2 u + q_1 \pm v^2)\]

When both $M$ and $N$ have parabolic points the singularity at each surface is of type $\widehat C_3^{*\pm}$ or type $\widehat C_3^{*}$.  We now prove this and derive the necessary conditions for each type. \hfill$\Box$

\medskip\noindent
{\bf Proof of Proposition \ref{C3-2-proof}}
 
 To show versality and in order to classify the type of $\widehat C_3^{*\pm}$ we need to keep track of higher degree terms in $q$.

Decompose $\Phi$ as a power series in $u,v$ by
 \[\Phi(u, v, \varepsilon. q_1, q_2, \tau) = \sum_{i, j} \phi_{ij}(\varepsilon, q_1, q_2, \tau)u^iv^j\]
 for some functions $\phi_{ij}$.

 Similarly to the previous case, substituting $u = \xi_1(u, v, q_1, q_2, \tau)$ and $v = \xi_2(u, v, q_1, q_2, \tau)$ for some functions $\xi_1$ and $\xi_2$, the higher degree terms
 can be removed, reducing the generating family $\Phi$ to the form

\[\tilde\Phi = -q_3 + \varepsilon(\tilde\phi_{00} + \tilde\phi_{10} u + \tilde\phi_{01} v  + \tilde\phi_{20}u^2   + u^3 \pm v^2)\]
for some functions $\tilde\phi_{ij}$ in variables $\varepsilon, q_1, q_2$ and $\tau$ which vanish at the origin.

Solving as a power series reveals that the necessary functions are 
\begin{eqnarray*}
\xi_1(u, v, \varepsilon, q_1, q_2, \tau) &=& \left(\frac{1}{\phi_{30}}\right)^\frac{1}{3}U -\frac{1}{3}\left(\frac{1}{\phi_{30}}\right)^\frac{5}{3}\phi_{40} U^2    - \frac{\phi_{30}\phi_{50}- \phi_{40}^2}{3\phi_{30}^3} U^3 +... \\
\xi_2(u, v, \varepsilon, q_1, q_2, \tau) &=& \nonumber \left(\frac{1}{\phi_{02}}\right)^\frac{1}{2}V -\frac{\phi_{03}}{2\phi_{02}^2}V^2 
-\frac{1}{2}\phi_{12}\left(\frac{1}{\phi_{02}}\right)^{\frac{3}{2}}UV
- \frac{\phi_{21}}{2\phi_{02}}U^2
- \frac{4\phi_{02}\phi_{04}  
- 5\phi_{03}^2}{8\phi_{02}^\frac{7}{2}} V^3 + ...
 \end{eqnarray*}
 
 Now a further change of coordinates $U = u_1 - \frac{\tilde\phi_{20}}{3}$ and   $V = v_1 - \frac{\tilde\phi_{01}}{2}$ gives 
 \[\widehat{\Phi} = -q_3 + \varepsilon(\widehat\phi_{00} + \widehat\phi_{10} u  + u^3 \pm v^2)\]
 for functions $\widehat\phi_{00}$ and $\widehat\phi_{10}$ not containing $u, v$ where
 
 \[\widehat\phi_{00}|_{\varepsilon=\tau=0} =  2 g_{02} q_2 + g_{11} q_1 + h.o.t.  \]
 
 \[\widehat\phi_{10}|_{\varepsilon=\tau=0} = -2 \frac{(f_{11}g_{20} - f_{20} g_{11})q_1}{f_{11}} - \frac{(f_{11}g_{11} - 4f_{20}g_{02})q_2}{f_{11}} + h.o.t.   \]
 
 Since the surface has a parabolic point, the linear terms in $q_2$ vanish and the necessary condition to be versally unfolded is that the terms
\[ \frac{1}{2} \frac{\partial \widehat\Phi}{\partial u \partial q_2^2} = 
-{\frac {24{{g_{02}}}^{3}{g_{30}}-12{{g_{02}}}^{2}{g_{11}}{
g_{21}}+6{{g_{11}}}^{2}{g_{12}}{g_{02}}-3{g_{03}}{{
g_{11}}}^{3}}{2{{g_{11}}}^{2}{g_{02}}}} \]
and $\frac{\partial^3 \widehat\Phi}{\partial u \partial \tau}$ are both nonzero, each of which provide independent conditions and so are avoided generically. 
  Further space-time contact transformations in parameters can reduce the generating family can be reduced to the normal form:

 \[\widehat C_3^{*\pm} : \mathcal{F} = -q_3 + \varepsilon(u^3 + u\varepsilon + \varepsilon + (\tau \pm q_2^2)u + q_1 \pm v^2),\]
where the $\pm u^2$ term is independent and does not affect the diffeomorphism type of the Centre Symmetry Set.

The sign of the term $\pm q_2^2$ is determined by the product of the derivatives
at $u=v=\varepsilon= q_1 = q_2=\tau=0$
\[ \frac{\partial^3 \widehat\Phi}{\partial u^3 } \frac{\partial^3 \widehat\Phi}{\partial u \partial q_2^2}  = -\left(f_{30}-{\frac {{f_{11}}{f_{21}}}{2{f_{02}}}}+{\frac {{{
f_{11}}}^{2}{f_{12}}}{4{{f_{02}}}^{2}}}-{\frac {{f_{03}}{{
f_{11}}}^{3}}{8{{f_{02}}}^{3}}} \right)\left(g_{30}-{\frac {{g_{11}}{g_{21}}}{2{g_{02}}}}+{\frac {{{
g_{11}}}^{2}{g_{12}}}{4{{g_{02}}}^{2}}}-{\frac {{g_{03}}{{
g_{11}}}^{3}}{8{{g_{02}}}^{3}}} \right)  \]
multiplied by a positive factor.

If this expression is negative then there is a lips bifurcation, and if positive then a beaks bifurcation on the binodal. The expression is zero if and only if one of the surfaces has a cusp of Gauss which is not true for this case.
\hfill$\Box$

\begin{remark} \label{rem:B3}A note about $\widehat{B_3}$.
{\rm
The condition here is, as Case 2 above, $g_{02}=0, g_{03} \neq 0$ and $M$ is not
parabolic at $(0,0,0)$. Analogously to (\ref{eq:linearterms}) and (\ref{eq:versalmatrix})
the singularity is versally unfolded if and only if the following matrix evaluated at the origin
$$
\begin{pmatrix}
\phi_1 & \phi_2 & \phi_3 \\
\frac{\partial \phi_1}{\partial \varepsilon} & \frac{\partial \phi_2}{\partial \varepsilon} & \frac{\partial \phi_3}{\partial \varepsilon}  \\
\frac{\partial^2 \phi_1}{\partial \varepsilon^2} & \frac{\partial ^2\phi_2}{\partial \varepsilon^2} & \frac{\partial^2 \phi_3}{\partial \varepsilon^2}
\end{pmatrix} =
\begin{pmatrix}
0 & 0 & -1 \\
g_{11} & 0 & 0   \\
2g_{11} + 2g_{12} & 6g_{03}  & 0
\end{pmatrix}$$
has nonzero determinant, which is clearly the case unless $g_{11}=0$.  If
this occurs (Case 2b), then to be versally unfolded as the more degenerate singularity  
\[
   \widehat B_3^{*} :  \mathcal{F} =
    -q_3 + \varepsilon(u^2 \pm v^2 \pm  \varepsilon^2 +  (\tau \pm q_1^2) \varepsilon + q_2)    
    \]
we would require the first two rows of the matrix to be independent. But $g_{11}=0$ prevents this
from happening, so $\widehat B_3^{*}$ does not occur in our geometrical context and instead this case results in the singularity $\widehat B_3^{**}$. 
 } 
\end{remark}

\section{Conclusions}\label{s:conclusions}

In this article we have applied the
techniques of modern singularity theory
to study the  curves and surfaces occurring
in the classical theory of thermodynamics applied to the mixing of
fluids, as derived by D.J.Korteweg 
and expounded by Sengers~\cite{Sengers}.
In fact to some extent we are responding
to a suggestion on page 85 of this book:
`It would be interesting to compare
Korteweg's method of continuous
deformation of surfaces with the
methodology of catastrophe theory'.
We have presented a complete list of
`normal forms' which describe the many
different `criminant surfaces' 
and `binodal curves' which arise
in this context, with proofs in
selected cases. The cases selected,
and the figures derived from them, are
of particular significance in
Korteweg's work.  One drawback of the
methods of singularity theory used here
is that certain geometrical information
is lost in passing to a normal form;
in particular the number of inflexions
on the binodal curves cannot
be predicted. For this we need to address
`dual' properties of the surfaces and
curves and this is the subject of
ongoing work.

\medskip\noindent
\centerline{\sc Acknowledgements}

\medskip\noindent
Graham Reeve is grateful to Liverpool Hope University for funding to attend Thermodynamics 2022 at The University of Bath and the 17th International Workshop on Real and Complex Singularities 2022 at The University of Sao Paulo where some of this work was presented.  Both authors would like to thank William Smith (The University of Guelph) and Mathias Brust (The University of Liverpool) for useful communications.

\bigskip\noindent
Peter Giblin, Department of Mathematical Sciences, The University of
Liverpool, Liverpool L69 7ZL, email pjgiblin@liv.ac.uk\\
Graham Reeve, School of Mathematics, Computer Science and Engineering, Liverpool
Hope University, Liverpool L16 9JD, email reeveg@hope.ac.uk


\begin{thebibliography}{99}
\bibitem{BGM} Thomas F.Banchoff, Terence Gaffney and Clint McCrory, {\em
Cusps of Gauss mappings}, Chapman and Hall/CRC Research Notes in Mathematics
(1982); web version by Daniel Dreibelbis at
https://www.emis.de/monographs/CGM/index.html
\bibitem{BGT}J.W.Bruce, P.J.Giblin and F.Tari, `Families of surfaces: height functions, Gauss maps and duals', in {\em Real and Complex Singularities}, W.L.Marar(ed.), Pitman Research Notes in Mathematics, Vol. 333 (1995), 148--178. 
\bibitem{Callen} Herbert B. Callen, {\em Thermodynamics and an Introduction to Thermostatistics}, Second Edition, John Wiley and Sons, 1985.
\bibitem{Denbigh}K. Denbigh, {\em The Principles of Chemical Equilibrium}, 4th ed., Cambridge University Press 1981.
\bibitem{Wojtek} Wojciech Domitrz, Miriam Manoel and Pedro de M. Rios, `The Wigner caustic on shell and singularities of odd functions', 
{\em Journal of Geometry and Physics} 71, 58-72.
\bibitem{Wojtek2} Wojciech Domitrz and  
Micha\l{} Zwierzyński, `The geometry of the Wigner caustic and a decomposition of a curve into parallel arcs' {\em Anal.Math.Phys} 12, Article 7, (2022).
\bibitem{DO}D. Dreibelbis and W. Olsen, `Structure and transitions of line bitangencies in a family of surface pairs',   {\it  Journal of Geometry} 113 (2022),
article 36 (29 pages).
\bibitem{GR}  Peter Giblin and Graham Reeve, `Equidistants for families of surfaces', {\em J. Singularities} 21 (2020), 97-118.
\bibitem{GR2} Peter J. Giblin and Graham Reeve, `Centre symmetry sets of families of plane curves'. {\em Demonstr Math} {\bf 48}, 167–192 (2015).
\bibitem{GZ1} P.J.Giblin and V.M.Zakalyukin, `Singularities of centre symmetry sets',
{\em Proc.\ London Math.\ Soc.\ } 90 (2005), 132--166.
\bibitem{GZ2} Peter J. Giblin and Vladimir M. Zakalyukin, `Recognition of centre
symmetry set singularities', {\em Geom.\ Dedicata} 130 (2007), 43--58.
\bibitem{K1} D.J. Korteweg,`Uber Faltenpunkte [On plait points]', {\em Sitzungsber. Akad. Wissensch. Wien, Math-Naturwissensch. Klasse}, Abt. 2A (1889) 1154--1191.
\bibitem{K2} D.J. Korteweg, `Vraagstukken [Problems] CXXXVII, CXXXVIII, CXXXIX', {\em Wiskundige opgaven
met oplossingen van het Genootschap E.O.A.K.A.T.B. [Mathematical exercises with solutions by
the society E.O.A.K.T.B.]} 4 (1890) 331--338. [The motto represented by the acronym signifies:
Untiring Labour Shall Overcome Everything.] Reprinted: Publication date 16 Apr 2012, Publisher: Nabu Press; Publication City/Country: Charleston SC, United States.
\bibitem{K3} D.J. Korteweg, `Sur les points de plissement', Arch. n\'{e}erl. 24 (1891a) 57--98. French
translation of \protect{\cite{K1}}.
\bibitem{K4} D.J. Korteweg, `La th\'{e}orie g\'{e}n\'{e}rale des plis', {\em Arch. n\'{e}erl.} 24 (1891b) 295--368.
\bibitem{K5}D.J. Korteweg, `Plaitpoints and corresponding plaits in the neighborhood of the sides of the $\psi$-surface
of Van der Waals', {\em Proc. Kon. Akad.} 5 (1903), 445--465. 
\bibitem{MSL} Gaetano Mangiapia, Roberto Sartorio and  Radu P. Lungu, 
`Plait point in ternary system with a miscibility gap: Analysis of the
procedures involved in its determination and development of novel
procedures', {\em Fluid Phase Equilibria} 409 (2016), 49--58.
\bibitem{math24} https://math24.net/van-der-waals-equation.html.
\bibitem{Meijer}Paul H.E.Meijer, `Theory of coexisting states: calculation of binodals', {\em Physica A: Statistical Mechanics
and its Applications}, 237 (1997) 31--44.
\bibitem{Olsen} William Edward Olsen, `Transitions in Line Bitangency Submanifolds for a One-Parameter Family of Immersion Pairs', University of North Florida Master's Thesis 2014
https://digitalcommons.unf.edu/etd/521/
\bibitem{ReeveThesis}Graham Reeve, {\em Singularities of Systems of Chords in Affine Space}, PhD thesis, University of Liverpool, 2012.
\bibitem{RZ1} Graham M. Reeve and Vladimir M. Zakalyukin, `Singularities of the Minkowski set and affine equidistants for a curve and a surface', {\em Topology and its Applications.} 159 (2), 555-561.
\bibitem{RZ} Graham M. Reeve and Vladimir M. Zakalyukin, `Singularities of the affine chord envelope for two surfaces in four-space', {\em Proc. Steklov Inst. of Math.} 277 (2012), 221--233.
\bibitem{RS} George Ruppeiner and Alex Seftas, `Thermodynamic Curvature of the Binary van der Waals Fluid', {\it Entropy}   22 (2020), 1208 (10 pages).
\bibitem{Sengers} Joanna Levelt Sengers, {\em How fluids unmix}, Koninklijke Nederlandse
Academie van Wetenschappen, Amsterdam 2002. ISBN 90-6984-357-9.
\bibitem{Ricardo} R. Uribe-Vargas, `A projective invariant for swallowtails and godrons, and global theorems on the flecnodal curve', {\em Moscow Math.\ J.\ }  6 (2006),731--768.
\bibitem{Varchenko} A.N.Varchenko, `Evolutions of convex hulls and phase transitions in thermodynamics', {\em J.\  Soviet Math.} 52 (1990), 3305--3325.
\end{thebibliography}
\end{document}